\documentclass[aps,amsmath,amssymb,twocolumn]{revtex4-1}

\usepackage{graphicx}
\usepackage{dcolumn}
\usepackage{bm}
\usepackage{color}

\begin{document}

\title{Magnetic-field control of near-field radiative heat transfer and the realization of highly 
tunable hyperbolic thermal emitters}

\author{E. Moncada-Villa$^{1}$}
\author{V. Fern\'andez-Hurtado$^{2}$}
\author{F. J. Garc\'{\i}a-Vidal$^{2}$}
\author{A. Garc\'{\i}a-Mart\'{\i}n$^{3}$}
\author{J. C. Cuevas$^2$}
\email{juancarlos.cuevas@uam.es}

\affiliation{$^{1}$Departamento de F\'{\i}sica, Universidad del Valle, AA 25360, Cali, Colombia}
\affiliation{$^2$Departamento de F\'{\i}sica Te\'orica de la Materia Condensada and Condensed Matter 
Physics Center (IFIMAC), Universidad Aut\'onoma de Madrid, E-28049 Madrid, Spain}
\affiliation{$^{3}$IMM-Instituto de Microelectr\'onica de Madrid (CNM-CSIC), Isaac Newton 8,
PTM, Tres Cantos, E-28760 Madrid, Spain}

\date{\today}

\begin{abstract}
We present a comprehensive theoretical study of the magnetic field dependence of the near-field
radiative heat transfer (NFRHT) between two parallel plates. We show that when the plates are 
made of doped semiconductors, the near-field thermal radiation can be severely affected by 
the application of a static magnetic field. We find that irrespective of its direction, the 
presence of a magnetic field reduces the radiative heat conductance, and dramatic reductions up
to 700\% can be found with fields of about 6 T at room temperature. We show that this 
striking behavior is due to the fact that the magnetic field radically changes the nature of the
NFRHT. The field not only affects the electromagnetic surface waves (both plasmons and 
phonon polaritons) that normally dominate the near-field radiation in doped semiconductors, but 
it also induces hyperbolic modes that progressively dominate the heat transfer as the field
increases. In particular, we show that when the field is perpendicular to the plates, the 
semiconductors become ideal hyperbolic near-field emitters. More importantly,
by changing the magnetic field, the system can be continuously tuned from a situation where
the surface waves dominate the heat transfer to a situation where hyperbolic modes completely
govern the near-field thermal radiation. We show that this high tunability can be achieved with 
accessible magnetic fields and very common materials like $n$-doped InSb or Si. Our study
paves the way for an active control of NFRHT and it opens the possibility to study unique 
hyperbolic thermal emitters without the need to resort to complicated metamaterials. 
\end{abstract}



\maketitle

\section{Introduction}

Radiative heat transfer is a topic with numerous fundamental and technological implications 
across disciplines \cite{Siegel2002}. Presently, the investigation of radiative heat transfer 
between closely spaced objects is receiving a great attention \cite{Zhang2007,Joulain2005,
Volokitin2007,Basu2009,Song2015a}. The basic challenges now are the understanding of the 
mechanisms governing thermal radiation at the nanoscale and the ability to control this
radiation for its use in novel applications. During a long time, it was believed that the maximum 
radiative heat transfer between two objects was set by the Stefan-Boltzmann law for black bodies. 
However, this is only true when objects are separated by distances larger than the thermal 
wavelength (9.6 $\mu$m at room temperature) and the radiative heat transfer is dominated 
by propagating waves. When two objects are brought in closer proximity, the thermal radiation 
is dominated by interference effects and, more importantly, by the near field emerging from 
the materials surfaces in the form of evanescent waves. This was first established theoretically 
by Polder and Van Hove in 1971 \cite{Polder1971} using Rytov's framework of fluctuational 
electrodynamics \cite{Rytov1953,Rytov1989}. These researchers predicted that the NFRHT could 
overcome the black body limit by several orders of magnitude, achieving what is referred to as 
super-Planckian thermal emission. Although this NFRHT enhancement was already hinted in 
several experiments in the late 1960's \cite{Hargreaves1969,Domoto1970}, its unambiguous 
confirmation came only in recent years \cite{Kittel2005,Narayanaswamy2008,Hu2008,Rousseau2009,
Shen2009,Ottens2011,Shen2012,Kralik2012,Zwol2012a,Zwol2012b,Guha2012,Shi2013,Worbes2013,
St-Gelais2014,Song2015b,Lim2015}. These experiments, in turn, have triggered off the hope that NFRHT 
may have an impact in different technologies such as heat-assisted magnetic recording 
\cite{Challener2009,Stipe2010}, thermal lithography \cite{Pendry1999}, scanning thermal 
microscopy \cite{Wilde2006,Kittel2008,Jones2013}, coherent thermal sources \cite{Carminati1999,
Greffet2002}, near-field based thermal management \cite{Otey2010,Rodriguez2011,Guha2012,Ben-Abdallah2014}, 
thermophotovoltaics \cite{Messina2013,Lenert2014} and other energy conversion devices \cite{Schwede2010}.

Presently, one of the central research lines in the field of radiative heat transfer is the 
search for materials where the NFRHT enhancement can be further increased. So far, the largest
enhancements have been experimentally reported in polar dielectrics \cite{Rousseau2009,
Shen2009,Song2015b}, where the near-field thermal radiation is dominated by the excitation of 
surface phonon polaritons (SPhPs) \cite{Mulet2002}. Similar enhancements have been predicted and 
observed in doped semiconductors due to surface plasmon polaritons (SPPs) \cite{Fu2006,Rousseau2009b,Shi2013,Lim2015}.
Recently, it has been predicted that hyperbolic metamaterials could behave as broadband super-Planckian
thermal emitters \cite{Nefedov2011,Biehs2012,Guo2012}. Hyperbolic metamaterials are a special class of highly 
anisotropic media that have hyperbolic dispersion. In particular, they are uniaxial materials for which one
of the principal components of either the permittivity or the permeability tensors is opposite in
sign to the other two principal components \cite{Smith2003}. Hyperbolic media have been mainly 
realized by means of hybrid metal-dielectric superlattices and metallic nanowires embedded in 
a dielectric host \cite{Poddubny2013,Shekhar2014}. It has been demonstrated that these metamaterials 
exhibit exotic optical properties such as negative refraction, subwavelength imaging and focusing,
and they can be used to do density of states engineering \cite{Poddubny2013,Shekhar2014}. 
In the context of radiative heat transfer, what makes these metamaterials so special is the 
fact that they can support electromagnetic modes that are evanescent in a vacuum gap, but they are
propagating inside the material. This leads to a broadband enhancement of transmission efficiency 
of the evanescent modes \cite{Biehs2012}. This special property has motivated a lot of theoretical work on
the use of hyperbolic metamaterials for NFRHT \cite{Biehs2013,Tschikin2013,Guo2013,Simovski2013,Liu2013,
Liu2014,Guo2014,Miller2014,Lang2014,Nefedov2014,Tschikin2015}. However, no experimental investigation 
of this issue has been reported so far, which is mainly due to the difficulties in handling these 
metamaterials. In this sense, it would be highly desirable to find much simpler realizations of 
hyperbolic thermal emitters and, ideally, with tunable properties.

Another key issue in the field of radiative heat transfer is the active control and modulation of 
NFRHT. In this respect, several proposals have been put forward recently. One of the them is 
based on the use of phase-change materials \cite{Zwol2011a,Zwol2011b}, where the change of phase 
leads to a significant change in the material dielectric function. These materials include an alloy called
AIST, where the phase change can be induced by applying an electric field \cite{Zwol2011b}, and VO$_2$, 
which undergoes a metal-insulator transition as a function of temperature \cite{Zwol2011a}. It has 
also been suggested that the NFRHT between chiral materials with magnetoelectric coupling can be tuned by
ultrafast optical pulses \cite{Cui2012}. Another proposal to tune the NFRHT is to use ferroelectric 
materials under an external electric field \cite{Huang2014}, although the predicted changes are rather
modest ($<17$\%). Let us also mention that very recently it has been proposed that the heat flux between
two semiconductors can be controlled by regulating the chemical potential of photons by means of an
external bias \cite{Chen2015}. So in short, although these proposals are certainly interesting, some of 
them are not easy to implement and others are either not very efficient or they are restricted to very 
specific materials. In this sense, the challenge remains to introduce strategies to actively control 
NFRHT in an easy and relatively universal way.

In this work we tackle and resolve some of the open problems described above by presenting an extensive theoretical
analysis of the influence of an external dc magnetic field in the radiative heat transfer between two 
parallel plates made of a variety of materials. We show that an applied magnetic field can indeed
largely affect the NFRHT in a broad class of materials, namely doped (polar and non-polar) semiconductors.
We find that, irrespective of its orientation, the magnetic field reduces the NFRHT with respect to
the zero-field case and we show that the reduction can be as large as 700\% for fields of about 6 T at 
room temperature. This effect originates from the fact that the magnetic field not only strongly modifies 
the surface waves that dominate the NFRHT in doped semiconductors (both SPhPs and SPPs), 
but it also generates broadband hyperbolic modes that tend to govern the heat transfer as the
field is increased. In particular, when the applied field is perpendicular to the plates surfaces,
the semiconductors behave as hyperbolic thermal emitters with highly tunable properties. By changing
the field magnitude one can continuously tune the system and realize situations where (i) surface waves
dominate the NFRHT, (ii) both surface waves and hyperbolic modes contribute significantly to the
near-field thermal radiation, and (iii) only hyperbolic modes contribute to the NFRHT and surface 
waves cease to exist. On the other hand, when the field is parallel to the surfaces the NFRHT is
nonmonotonic as a function of the magnetic field. For moderate fields, surface waves and hyperbolic 
modes coexisting, while for high fields the NFRHT is largely dominated by hyperbolic modes. We emphasize 
that all these striking predictions are amenable to measurements and do not require the use of any complicated
metamaterial. Thus, our work offers a simple strategy to actively control NFRHT in a broad variety
of materials and it also provides a very appealing recipe to realize hyperbolic materials and, in
particular, hyperbolic thermal emitters with highly tunable properties. 

The remainder of this paper is structured as follows. Section \ref{sec-II} describes the system
under study and the general formalism for the description of NFRHT in the presence of a magnetic
field. We then turn in Sec. \ref{sec-III} to the application of the general results to the case 
of $n$-doped InSb as an example of a polar semiconductor. We discuss in this section both the 
results for different magnetic field orientations and the realization of highly tunable 
hyperbolic thermal emitters. Section \ref{sec-III} is devoted to the case of Si as an 
example of non-polar semiconductor. Section IV summarizes our main results and discusses
future directions. Finally, four appendixes contain the technical details of the general 
formalism and some additional calculations that support the claims in this paper.

\section{Radiative heat transfer in the presence of a magnetic field: General formalism} \label{sec-II}

Our main goal is to compute the radiative heat transfer in the presence of an external dc magnetic
field within the framework of fluctuational electrodynamics \cite{Rytov1953,Rytov1989}. For simplicity, 
we shall concentrate here in the heat exchanged between two infinite parallel plates made of arbitrary 
non-magnetic materials and that are separated by a vacuum gap of width $d$, see Fig.~\ref{scheme}(a). 
The magnetic field can point in any direction and following Fig.~\ref{scheme}(a), we shall refer
to the left plate as medium 1, the vacuum gap as medium 2, and the right plate as medium 3. 

\begin{figure}[t]
\begin{center} \includegraphics[width=0.85\columnwidth,clip]{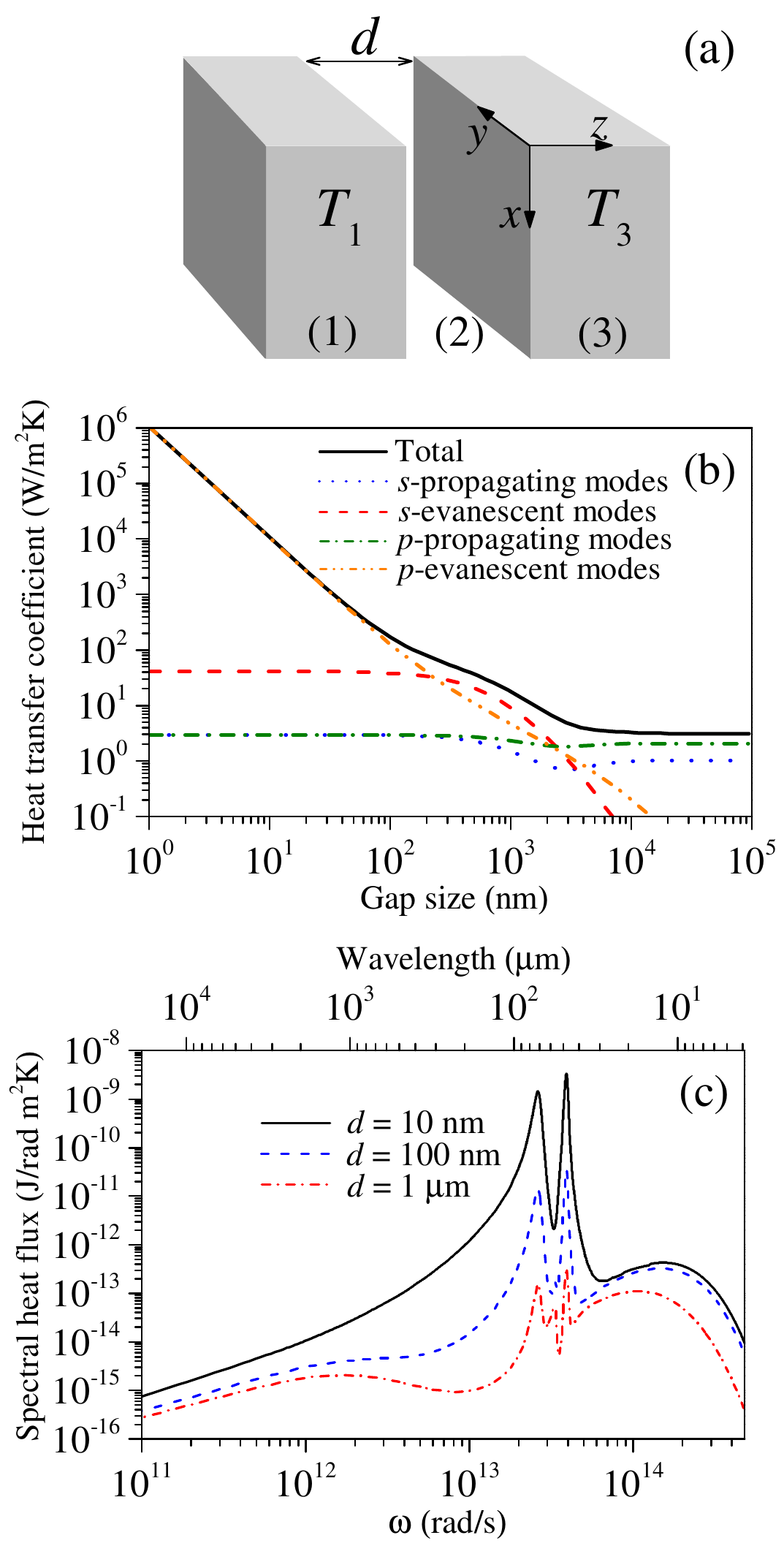} \end{center}
\caption{(a) Schematic representation of the system under study: two parallel plates at temperatures
$T_1$ and $T_3$ separated by a vacuum gap of width $d$. (b) Heat transfer coefficient of $n$-doped
InSb as a function of the gap at zero magnetic field. We show the total result and the individual
contributions of $s$- and $p$-polarized waves (both propagating and evanescent). (c) The corresponding 
zero-field spectral heat flux as a function of frequency (and wavelength) for three different gaps.}
\label{scheme}
\end{figure}

When a magnetic field is applied to any object, it results in an optical anisotropy that can be 
described by the following general permittivity tensor \cite{Zvezdin1997}
\begin{equation}
\label{perm-tensor}
\hat \epsilon = \left( \begin{array}{ccc}
\epsilon_{xx} & \epsilon_{xy} & \epsilon_{xz} \\
\epsilon_{yx} & \epsilon_{yy} & \epsilon_{yz} \\
\epsilon_{zx} & \epsilon_{zy} & \epsilon_{zz} \end{array} \right) ,
\end{equation}
where according to Fig.~\ref{scheme}(a), $x$ and $y$ lie in the interface planes and $z$
corresponds to the surface normal. The components of the permittivity tensor depend on the applied 
magnetic field, as we shall specify below, and on the frequency (local approximation). Let us recall 
that the off-diagonal elements in Eq.~(\ref{perm-tensor}) are responsible for all the well-known 
magneto-optical effects (Faraday effect, Kerr effects, etc.) \cite{Zvezdin1997}. Thus, our problem 
is to compute the radiative heat transfer between two anisotropic parallel plates. This generic 
problem has been addressed by Biehs \emph{et al.}\ \cite{Biehs2011} and we just recall here the 
central result. The net power per unit of area exchanged between the parallel plates is given by  
the following Landauer-like expression \cite{Biehs2011}
\begin{equation}
\label{eq-net-Q}
Q = \int^{\infty}_{0} \frac{d \omega}{2\pi} \left[ \Theta_1(\omega) - \Theta_3(\omega) 
\right] \int \frac{d{\bf k}}{(2\pi)^2} \tau(\omega,{\bf k},d) ,
\end{equation}
where $\Theta_i(\omega) = \hbar \omega/ [\exp(\hbar \omega / k_{\rm B}T_i) -1]$, $T_i$ is the absolute 
temperature of the layer $i$, $\omega$ is the radiation frequency, ${\bf k} = (k_x,k_y)$ is the wave 
vector parallel to the surface planes, and $\tau(\omega,{\bf k},d)$ is the total transmission 
probability of the electromagnetic waves. Notice that the second integral in Eq.~(\ref{eq-net-Q})
is carried out over all possible directions of ${\bf k}$ and it includes the contribution of both
propagating waves with $k < \omega/c$ and evanescent waves with $k> \omega/c$, where $k$ is the
magnitude of ${\bf k}$ and $c$ is the velocity of light in vacuum. The transmission coefficient
$\tau(\omega,{\bf k},d)$ can be expressed as \cite{Biehs2011}
\begin{eqnarray}
\label{eq-trans-man}
\tau(\omega,{\bf k},d) = \hspace{7cm} & & \\ \left\{ \begin{array}{ll}
\mbox{Tr} \left\{ [\hat 1 - \hat {\cal R}_{21} \hat {\cal R}^{\dagger}_{21} ] \hat {\cal D}^{\dagger}
[\hat 1 - \hat {\cal R}^{\dagger}_{23} \hat {\cal R}_{23} ] \hat {\cal D} \right\}, & k < \omega/c \\
\mbox{Tr} \left\{ [\hat {\cal R}_{21} - \hat {\cal R}^{\dagger}_{21} ] \hat {\cal D}^{\dagger}
[\hat {\cal R}^{\dagger}_{23} - \hat {\cal R}_{23} ] \hat {\cal D} \right\} e^{-2|q_2|d}, & k > \omega/c
\end{array} \right. , & & \nonumber
\end{eqnarray}
where $q_2 = \sqrt{\omega^2/c^2 - k^2}$ is the $z$-component of the wave vector in the vacuum gap and the
$2 \times 2$ matrices $\hat {\cal R}_{ij}$ are the reflections matrices characterizing the two interfaces.
These matrices have the following generic structure
\begin{equation}
\label{refl-mat}
\hat {\cal R}_{ij} = \left( \begin{array}{cc} r^{s,s}_{ij} & r^{s,p}_{ij} \\ 
r^{p,s}_{ij} & r^{p,p}_{ij} \end{array} \right) ,
\end{equation}
where $r^{\alpha, \beta}_{ij}$ with $\alpha,\beta =s,p$ is the reflection amplitude for the 
scattering of an incoming $\alpha$-polarized plane wave into an outgoing $\beta$-polarized wave.
Finally, the $2\times 2$ matrix $\hat {\cal D}$ appearing in Eq.~(\ref{eq-trans-man}) is defined
as
\begin{equation}
\hat {\cal D} = [ \hat 1 - \hat {\cal R}_{21} \hat {\cal R}_{23} e^{2iq_2d} ]^{-1}.
\end{equation}
Notice that this matrix describes the usual Fabry-P\'erot-like denominator resulting from the 
multiple scattering between the two interfaces.

In Appendixes A and B we provide an alternative derivation of the central result of Eq.~(\ref{eq-trans-man})
that emphasizes the non-reciprocal nature of our problem. More importantly, we show explicitly how 
the different reflection matrices appearing in Eq.~(\ref{eq-trans-man}) can be computed within a 
scattering-matrix approach for anisotropic multilayer systems. This approach provides, in turn,
a natural framework to analyze different issues that will be crucial later on such as the nature
of the electromagnetic modes responsible for the heat transfer.

The result of Eqs.~(\ref{eq-net-Q}) and (\ref{eq-trans-man}) reduces to the
well-known result for isotropic media first derived by Polder and Van Hove \cite{Polder1971}. In that
case, the reflections matrices of Eq.~(\ref{refl-mat}) are diagonal and the non-vanishing elements are 
given by
\begin{eqnarray}
\label{eq-rs-iso}
r^{s,s}_{ij} & = & \frac{q_i - q_j}{q_i + q_j} \\
\label{eq-rp-iso}
r^{p,p}_{ij} & = & \frac{\epsilon_j q_i - \epsilon_i q_j}{\epsilon_j q_i + \epsilon_i q_j} ,
\end{eqnarray}
where $q_i = \sqrt{\epsilon_i \omega^2 /c^2 - k^2}$ is the transverse or $z$-component of the wave vector
in layer $i$ and $\epsilon_i(\omega)$ is the corresponding dielectric constant. Thus, the total transmission
can be written as $\tau(\omega,{\bf k},d) = \tau_s(\omega,{\bf k},d) + \tau_p(\omega,{\bf k},d)$, where 
the contributions of $s$- and $p$-polarized waves are given by
\begin{eqnarray}
\label{eq-trans-iso}
\tau_{\alpha=s,p}(\omega,{\bf k},d) = \hspace{6cm} && \\ \left\{ \begin{array}{ll}
(1 - |r^{\alpha,\alpha}_{21}|^2) ( 1 - |r^{\alpha,\alpha}_{23}|^2)/|D_{\alpha}|^2 , \;\; k < \omega/c \\
4 \mbox{Im} \{ r^{\alpha,\alpha}_{21} \} \mbox{Im} \{ r^{\alpha,\alpha}_{23} \} e^{-2|q_2|d} /|D_{\alpha}|^2
, \;\; k > \omega/c \end{array} \right. , && \nonumber
\end{eqnarray}
where $D_{\alpha} = 1 - r^{\alpha,\alpha}_{21} r^{\alpha,\alpha}_{23} e^{2iq_2d}$.
Throughout this work we focus on the analysis of the radiative linear heat conductance per unit
of area, $h$, which is referred to as the heat transfer coefficient. This coefficient is given by
\begin{equation}
h(T,d) = \lim_{\Delta T \rightarrow 0^+} \frac{Q(T_1=T+\Delta T,T_3=T,d)}{\Delta T} , 
\end{equation}
where $T$ is the absolute temperature that we assume equal to 300 K throughout this work.
Additionally, we define the spectral heat flux as the heat transfer coefficient per unit of frequency.
In the following sections, we apply the general results presented here to different materials and 
magnetic field configurations.

\section{Polar semiconductors: $\mbox{InSb}$} \label{sec-III}

The first obvious question to be answered is: In which materials can a magnetic field modify
the NFRHT? Since the thermal radiation of an object is primarily determined by its dielectric
function, we need materials in which this function can be modified by an external magnetic field, 
that is we need magneto-optical (MO) materials. Focusing on room temperature experiments, the MO
activity must be exhibited in the mid-infrared. Thus, doped semiconductors, where the MO activity
is due to conduction electrons, are ideal candidates \cite{Kushwaha2001}. In these materials, 
one can play with the doping level to tune the plasma frequency to values comparable to the 
cyclotron frequency at experimentally achievable magnetic fields, which is an important requirement 
to have sizable magnetic-induced effects in the NFRHT (see discussion below). Moreover, in
semiconductors the NFRHT in the absence of field is typically dominated by surface electromagnetic 
waves (both SPhPs and SPPs), which in turn are known to be strongly influenced by an external
magnetic field \cite{Kushwaha2001,Wallis1982}. Thus, it seems natural to expect a 
magnetic-field modulation of NFRHT in semiconductors.

There is a variety of semiconductors that we could choose to illustrate our predictions.
In this section we focus on InSb for several reasons. First, it is a polar semiconductor
where the NFRHT in the absence of field is dominated by two different types of surface waves
(SPhPs and SPPs), which allows us to study a very rich phenomenology. Second, InSb has a small 
effective mass, which enables to tune the cyclotron frequency to values comparable to those of 
the plasma frequency with moderate fields. Finally, InSb has been the most widely studied material in the 
context of magnetoplasmons and coupled magnetoplasmons-surface phonon polaritons. Thus, the
magnetic field effect in the surface waves has been very well characterized experimentally
\cite{Palik1973,Hartstein1975,Palik1976,Remer1984}. 

\subsection{Perpendicular magnetic field: The realization of hyperbolic near-field
thermal emitters}

Let us first discuss the radiative heat transfer between two identical plates made of $n$-doped InSb 
when the magnetic field is perpendicular to the plate surfaces, \emph{i.e.}\ ${\bf H} = H_z \hat {\bf z}$, 
see Fig.~\ref{scheme}(a). In this case, the permittivity tensor of InSb adopts the following
form \cite{Palik1976}
\begin{equation}
\label{perm-tensor-Hz}
\hat \epsilon(H) = \left( \begin{array}{ccc} \epsilon_1(H) & -i\epsilon_2(H) & 0 \\
i \epsilon_2(H) & \epsilon_1(H) & 0 \\ 0 & 0 & \epsilon_3 \end{array} \right) ,
\end{equation}
where
\begin{eqnarray}
\epsilon_1(H) & = & \epsilon_{\infty} \left( 1 + \frac{\omega^2_L - \omega^2_T}{\omega^2_T - 
\omega^2 - i \Gamma \omega} + \frac{\omega^2_p (\omega + i \gamma)}{\omega [\omega^2_c -
(\omega + i \gamma)^2]} \right) , \nonumber \\
\label{eq-epsilons}
\epsilon_2(H) & = & \frac{\epsilon_{\infty} \omega^2_p \omega_c}{\omega [(\omega + i \gamma)^2 -
\omega^2_c]} , \\
\epsilon_3 & = & \epsilon_{\infty} \left( 1 + \frac{\omega^2_L - \omega^2_T}{\omega^2_T -
\omega^2 - i \Gamma \omega} - \frac{\omega^2_p}{\omega (\omega + i \gamma)} \right) . \nonumber
\end{eqnarray}
Here, $\epsilon_{\infty}$ is the high-frequency dielectric constant, $\omega_L$ is the longitudinal
optical phonon frequency, $\omega_T$ is the transverse optical phonon frequency, $\omega^2_p =
ne^2/(m^{\ast} \epsilon_0 \epsilon_{\infty})$ defines the plasma frequency of free carriers of density 
$n$ and effective mass $m^{\ast}$, $\Gamma$ is the phonon damping constant, and $\gamma$ is the 
free-carrier damping constant. Finally, the magnetic field enters in these expressions via the cyclotron 
frequency $\omega_c = eH/m^{\ast}$. The important features of the previous expressions are:
(i) the magnetic field induces an optical anisotropy (via the modification of the diagonal elements
and the introduction of off-diagonal ones), (ii) there are two major contributions to the diagonal
components of the dielectric tensor: optical phonons and free carriers, and (iii) the MO activity
is introduced via the free carriers, which illustrates the need to deal with doped semiconductors.
In what follows we shall concentrate in a particular case taken from Ref.~[\onlinecite{Palik1976}],
where $\epsilon_{\infty} = 15.7$, $\omega_L = 3.62 \times 10^{13}$ rad/s, $\omega_T = 3.39\times 10^{13}$ 
rad/s, $\Gamma = 5.65 \times 10^{11}$ rad/s, $\gamma = 3.39 \times 10^{12}$ rad/s, $n = 1.07 \times 10^{17}$ 
cm$^{-3}$, $m^{\ast}/m = 0.022$, and $\omega_p = 3.14 \times 10^{13}$ rad/s. As a reference, let us
say that with these parameters $\omega_c = 8.02 \times 10^{12}$ rad/s for a field of 1 T.
Let us point out that in this configuration, and due to the structure of the permittivity tensor, the 
transmission coefficient appearing in Eq.~(\ref{eq-net-Q}) only depends on the magnitude of the 
parallel wave vector, which considerably simplifies the calculation of the radiative heat transfer.

Let us now briefly review the expectations for the heat transfer in the absence of magnetic field. 
As we show in Fig.~\ref{scheme}(b), the heat transfer coefficient features a large near-field 
enhancement for gaps below 1 $\mu$m. For $d < 100$ nm this enhancement is largely dominated by 
$p$-polarized evanescent waves and the heat transfer coefficient increases as $1/d^2$ as the gap 
decreases, which are two clear signatures of a situation where the heat transfer is dominated by 
surface electromagnetic waves. This can be further confirmed with the analysis of the spectral heat 
flux, see Fig.~\ref{scheme}(c), which in the near-field regime is dominated by two narrow peaks that 
can be associated to SPPs (low-frequency peak) and SPhPs (high-frequency peak), as it will become 
evident below. Thus, the case of InSb constitutes an interesting example where two types of surface
waves contribute significantly to the NFRHT. Let us now see how these results are modified in the 
presence of a magnetic field.

\begin{figure}[t]
\begin{center} \includegraphics[width=0.85\columnwidth,clip]{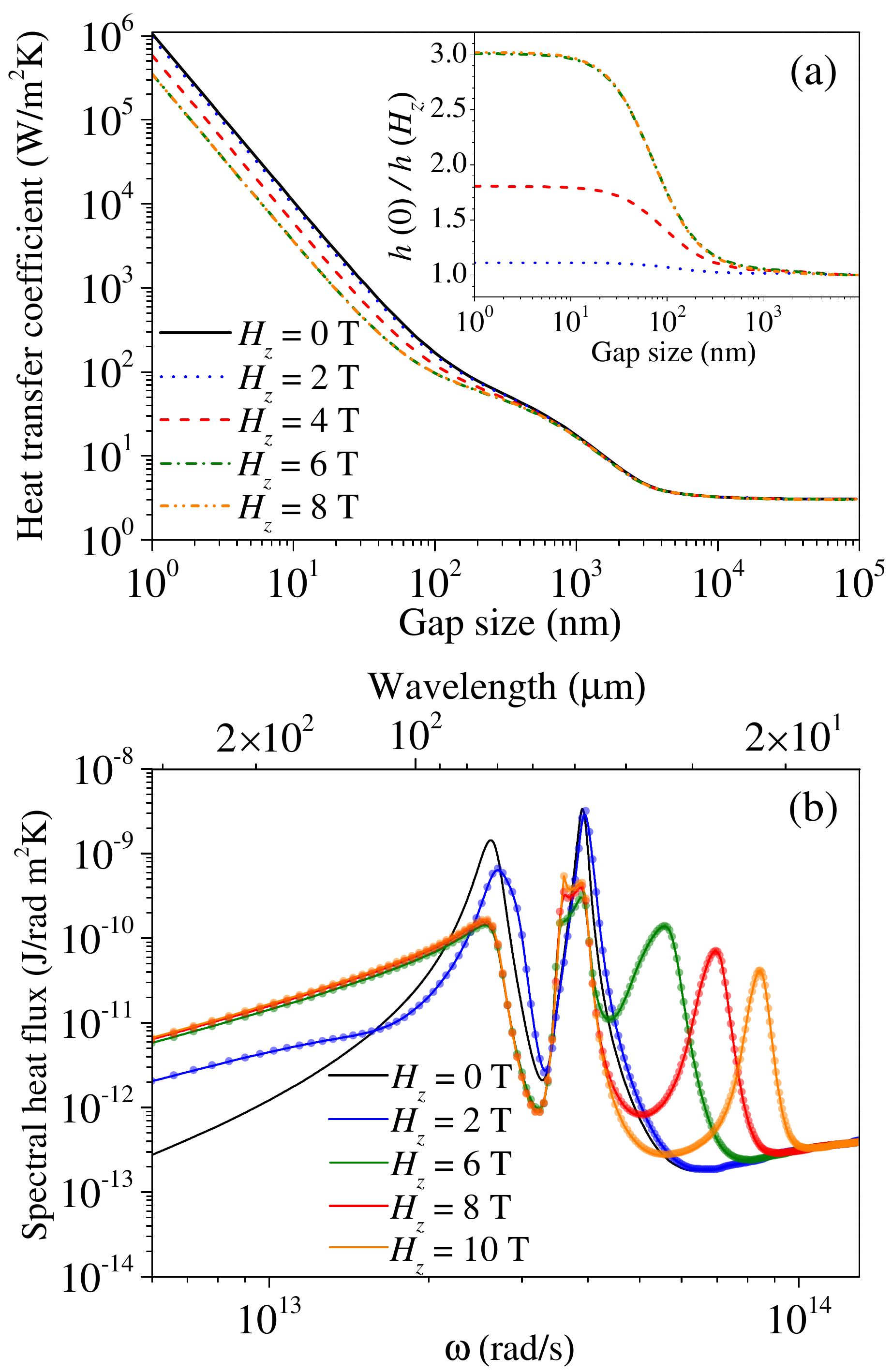} \end{center}
\caption{(a) Heat transfer coefficient for $n$-doped InSb as a function of the gap for different
values of the magnetic field perpendicular to the plate surfaces. The inset shows
the ratio between the zero-field coefficient and the coefficient for different values of the
field in the near-field region. (b) The corresponding spectral heat flux as a function of
the frequency (and wavelength) for a gap of $d=10$ nm and different values of the perpendicular
field. The solid lines correspond to the exact calculation and the circles to the uniaxial approximation
where the off-diagonal terms of the permittivity tensor are assumed to be zero.}
\label{fig-Hz}
\end{figure}

In Fig.~\ref{fig-Hz}(a) we show the heat transfer coefficient as a function of the gap size for 
different values of the perpendicular magnetic field. There are three salient features: (i) the 
far-field heat transfer is fairly independent of the magnetic field, (ii) in the near-field regime (below 
300 nm) the magnetic field suppresses the heat transfer by up to a factor of 3 (see inset), and 
(iii) by increasing the field, the heat transfer coefficient tends to saturate at around 6 T, although
it is slightly reduced upon further increasing the field above 10 T (not shown here). The strong 
modification of heat transfer due to the magnetic field is even more apparent in the spectral heat 
flux. As one can see in Fig.~\ref{fig-Hz}(b), the magnetic field not only distorts and reduces the 
height of the peaks related to the surface waves, but it also generates a new peak that shifts to 
higher frequencies as the field increases. This additional peak appears at the cyclotron 
frequency and its presence illustrates the high tunability that can be achieved. Notice, for instance, 
that for a field of 6 T the thermal emission at the cyclotron frequency is increased by almost 3 orders 
of magnitude with respect to the zero-field case. 

\begin{figure*}[t]
\includegraphics*[width=\textwidth,clip]{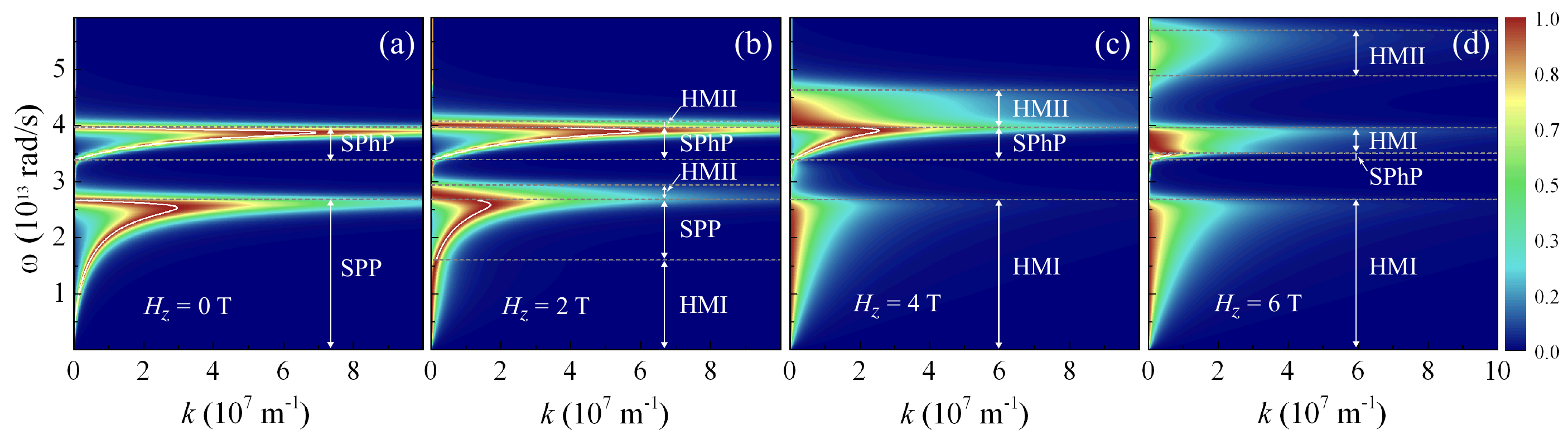}
\caption{The transmission coefficient for $p$-polarized waves as a function of the magnitude
of the parallel wave vector and frequency for InSb and a gap of $d=10$ nm. The different panels
correspond to different values of the magnetic field that is perpendicular to the surfaces. 
The horizontal dashed lines separate the regions where transmission is dominated by surface waves 
(SPPs and SPhPs) or hyperbolic modes of type I and II (HMI and HMII). The white solid lines 
correspond to the analytical dispersion relation of the surface waves of Eq.~(\ref{dr-Hz}).}
\label{fig-trans-Hz}
\end{figure*}

To shed more light on these results it is convenient to examine the transmission of the $p$-polarized
waves, which can be shown to dominate the heat transfer for any field. We present in Fig.~\ref{fig-trans-Hz} 
this transmission as a function of the magnitude of the parallel wave vector, $k$, and the frequency for 
a gap $d=10$ nm and different values of the magnetic field. As one can see, at low fields the transmission
maxima are located around a restricted area of $k$ and $\omega$, clearly indicating that surface
waves dominate the NFRHT. Notice also that their dispersion relation is modified by the field, see
Fig.~\ref{fig-trans-Hz}(b). By increasing the field, those areas are progressively replaced by 
areas where the maximum transmission is reached for a broad range of $k$-values and finally, the
surface waves are restricted to the reststrahlen band $\omega_T < \omega < \omega_L$ for the 
highest fields, see Fig.~\ref{fig-trans-Hz}(d). What is the nature of these magnetic-field-induced modes?  

To answer this question and explain all the results just described, it is important to realize that
the off-diagonal elements of the permittivity tensor do not play a major role in this configuration.
This is illustrated in Fig.~\ref{fig-Hz}(b) where we show that the approximation consisting in setting
$\epsilon_2 = 0$ in Eq.~(\ref{perm-tensor-Hz}) reproduces very accurately the exact results for the 
spectral heat flux for arbitrary magnetic fields. This means that the polarization conversion is 
irrelevant and the plates effectively behave as uniaxial media where their permittivity tensors 
are diagonal: $\hat \epsilon=$ \emph{diag}$[\epsilon_{xx},\epsilon_{xx},\epsilon_{zz}]$, where $\epsilon_{xx} = 
\epsilon_1$ and $\epsilon_{zz} = \epsilon_3$. Within this approximation, which hereafter we refer to as 
uniaxial approximation, it is easy to compute the dispersion relation of the surface electromagnetic
modes in our geometry (see Appendix D). In the electrostatic limit $k \gg \omega/c$, the dispersion
relation of these cavity modes is given by
\begin{equation}
\label{dr-Hz}
k_{\rm SW} = \frac{1}{d} \ln \left( \pm \frac{\epsilon_{xx} - \sqrt{\epsilon_{xx}/\epsilon_{zz}}} 
{\epsilon_{xx} + \sqrt{\epsilon_{xx}/\epsilon_{zz}}} \right) ,
\end{equation}
with the additional constraint that both $\epsilon_{xx}$ and $\epsilon_{zz}$ must be negative. In 
the zero-field limit, this expression reduces to the known result for cavity surface modes in isotropic 
materials \cite{Song2015b}. As we show in Fig.~\ref{fig-trans-Hz}, see white solid lines, the 
dispersion relation of Eq.~(\ref{dr-Hz}) 
nicely reproduces the structure of the transmission maxima in those frequency regions in which surfaces
waves are allowed ($\epsilon_{xx},\epsilon_{zz} < 0$). It is worth stressing that this dispersion 
relation describes in a unified manner both the SPPs that appear below the reststrahlen band and 
the SPhPs due to the optical phonons. More importantly, this dispersion relation tells us that the 
magnetic field reduces the parallel wave vector of the surface waves and restricts the frequency 
region where they exist. Indeed, at high fields the SPPs disappear, while the SPhPs are restricted
to the reststrahlen band, Fig.~\ref{fig-trans-Hz}(d). These two effects are actually the cause of the 
reduction of the NFRHT in the presence of a magnetic field. But what about the other modes that appear by 
increasing the field? Their nature can also be understood within the uniaxial approximation. As we show 
in Appendix C, the allowed values for the transverse component of the wave vector inside these 
uniaxial materials are given by $q_{\rm o} = \sqrt{\epsilon_{xx} \omega^2/c^2 - k^2}$ for ordinary 
waves and $q_{\rm e} = \sqrt{\epsilon_{xx} \omega^2/c^2 - k^2 \epsilon_{xx}/\epsilon_{zz}}$ for 
extraordinary waves. The dispersion of the extraordinary waves can be rewritten as
\begin{equation}
\label{dr-HM}
\frac{k^2_x + k^2_y}{\epsilon_{zz}} + \frac{q^2_{\rm e}}{\epsilon_{xx}} = \frac{\omega^2}{c^2} ,
\end{equation}
a dispersion that becomes hyperbolic when $\epsilon_{xx}$ and $\epsilon_{zz}$ have opposite signs 
\cite{Smith2003}. This is exactly what happens in our case in certain frequency regions at finite 
field. This illustrated in Fig.~\ref{fig-trans-Hz}(b-d), where we have indicated the hyperbolic regions 
defined by the condition $\epsilon_{xx} \epsilon_{zz} < 0$. Notice that those regions correspond 
exactly to the areas where the transmission reaches its maximum for a broad range of $k$-values. This 
fact shows unambiguously that our InSb plates effectively behave as hyperbolic materials. More 
importantly, and as it is evident from Fig.~\ref{fig-trans-Hz}, we can easily modify the hyperbolic 
regions by changing the field. Thus, we can change from situations where the hyperbolic modes (HMs) 
coexist with both types of surface waves to situations where the HMs dominate the NFRHT, which 
is what occurs at high fields, see Fig.~\ref{fig-trans-Hz}(d). Moreover, contrary to what happens 
in most hybrid hyperbolic metamaterials, we can have in a single material HMs of type I (HMI), where 
$\epsilon_{xx} > 0$ and $\epsilon_{zz} < 0$, and HMs of type II (HMII), where $\epsilon_{xx} < 0$ 
and $\epsilon_{zz} > 0$, see Fig.~\ref{fig-trans-Hz}(b-d).

Let us recall that what makes HMs so special in the context of NFRHT is the fact that, as it is
evident from their dispersion relation, they are evanescent in the vacuum gap and propagating inside 
the hyperbolic material for $k > \omega/c$ (HMI) or $k > \sqrt{|\epsilon_{zz}|} \omega/c$ (HMII).
Thus, they are a special kind of frustrated internal reflection modes that exhibit a very high 
transmission over a broad range of $k$-values that correspond to evanescent waves in the vacuum 
gap \cite{Biehs2012}. As shown in Ref.~[\onlinecite{Biehs2012}], 
the number of HMs that contributes to the NFRHT is solely determined by the intrinsic cutoff in the 
transmission, which has the form $\tau(\omega,k) \propto \exp(-2kd)$ for $k \gg \omega/c$. From 
this condition it follows that the heat flux due to HMs scales as $1/d^2$ for small gaps, as 
the contribution of surface waves. This explains why the appearance of HMs as the field 
increases does not modify the parametric dependence of the NFRHT with the gap size. Notice,
however, that in spite of the high transmission of these HMs, their appearance does not enhance
the NFRHT because they replace surface waves that possess even larger $k$-values (notice that
the conditions of HMs and surface waves are mutually excluding). Thus, we can conclude that the NFRHT
reduction induced by the magnetic field is due to both the modification of the surface waves and
their replacement by HMs that, in spite of their propagating nature inside the material, turn
out to be less efficient transferring the radiative heat in the near-field region than the surface
waves.

Let us point out that within the uniaxial approximation, the heat transfer can be obtained in
a semi-analytical form. In this case, the transmission coefficient is given by the isotropic
result of Eq.~(\ref{eq-trans-iso}), where the reflections coefficients adopt now the form
\begin{eqnarray}
\label{eq-rs-Hz}
r^{s,s}_{21} = r^{s,s}_{23} & = & \frac{q_2 - q_{\rm o}}{q_2 + q_{\rm o}} \\
\label{eq-rp-Hz}
r^{p,p}_{21} = r^{p,p}_{23} & = & \frac{\epsilon_{xx} q_2 - q_{\rm e}}{\epsilon_{xx} q_2 + q_{\rm e}} .
\end{eqnarray}

The uniaxial approximation is also useful to understand the high field behavior of the NFRHT. The
tendency to saturate the thermal radiation as the field increases is due to the to the fact that 
the cyclotron frequency becomes larger than the plasma frequency and the last term in the expression
of $\epsilon_{xx}=\epsilon_1$, see Eq.~(\ref{eq-epsilons}), progressively becomes more irrelevant.
Thus, the permittivity tensor becomes field-independent and the heat transfer is simply given by 
the result for uniaxial media, where $\epsilon_{zz}=\epsilon_3$ has the form in Eq.~(\ref{eq-epsilons}),
but $\epsilon_{xx}=\epsilon_1$ does not contain the last term in the first expression of Eq.~(\ref{eq-epsilons}).
We find that the strict saturation of the NFRHT occurs at around 20 T and there is an intermediate regime,
between 6 and 20 T, in which the near-field thermal radiation slightly increases upon increasing the field
(not shown here), leading to a nonmonotonic behavior. This behavior is due to an increase in the
efficiency of the HMs that dominate the NFRHT in this high-field regime.

\begin{figure}[t]
\begin{center} \includegraphics[width=0.86\columnwidth,clip]{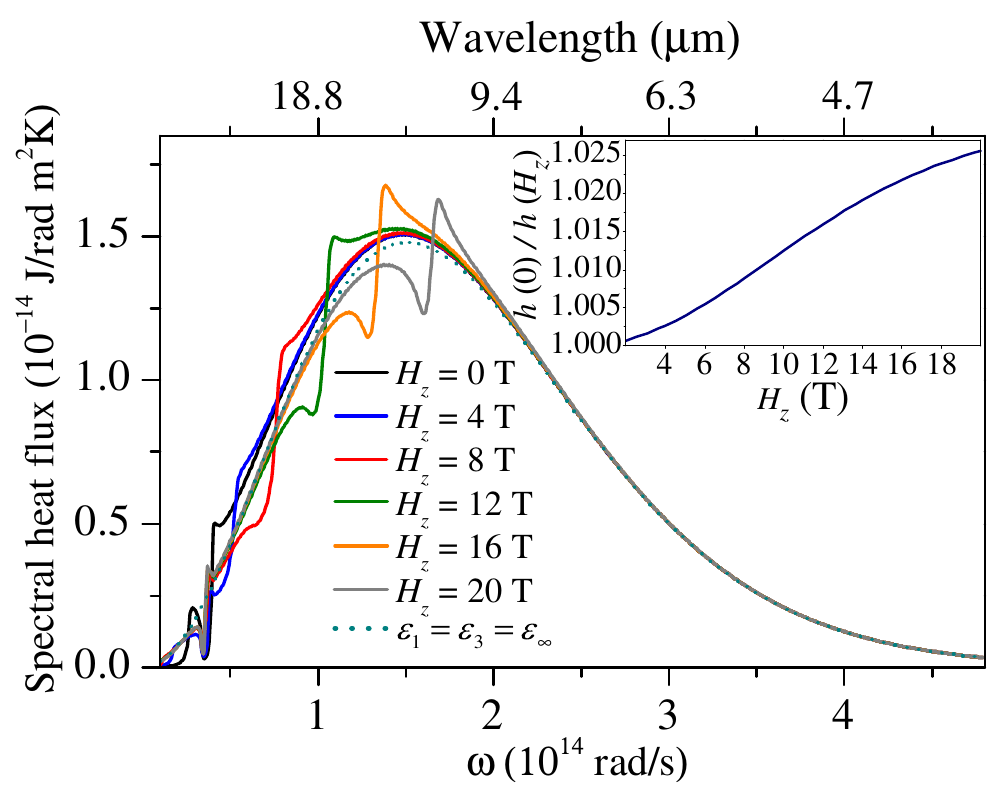} \end{center}
\caption{Far-field spectral heat flux for InSb as a function of the frequency (and wavelength) for
different values of the perpendicular field. These spectra have been computed for a gap $d=1$ m. The dotted 
line corresponds to the result for plates made of a dielectric with a frequency-independent dielectric
constant equal to $\epsilon_{\infty}$. The inset shows the corresponding ratio between the zero-field heat
transfer coefficient and the coefficient for different values of the field.}
\label{fig-far-field}
\end{figure}

To conclude this subsection, let us explain why the far-field heat transfer is rather insensitive to the 
magnetic field. For gaps much larger than the thermal wavelength ($9.6\; \mu$m), the heat transfer is 
dominated by propagating waves and, as we show in Fig.~\ref{fig-far-field}, the spectral heat flux in the 
absence of field exhibits a broad spectrum with a peak at around $1.5\times 10^{14}$ rad/s. Indeed, the spectrum 
is very similar to that of a dielectric with a frequency-independent dielectric constant $\hat \epsilon = \epsilon_{\infty} 
\hat 1$, see dotted line in Fig.~\ref{fig-far-field}. As we illustrate in that figure, the presence of a magnetic 
field only modifies this spectrum in a significant way in a small region around the cyclotron frequency. This 
fact leads to a tiny modification of the heat transfer upon the application of an external field. As it can
be seen in the inset of Fig.~\ref{fig-far-field}, the magnetic field reduces the far-field heat transfer 
coefficient as the magnetic field increases, but this reduction is quite modest and, for instance, it amounts 
to only 2.5\% at a very high field of 20 T. 

\subsection{Parallel magnetic field}

Let us now turn to the case in which the magnetic field is parallel to the plate surfaces. For concreteness,
we consider that the field is applied along the $x$-axis, ${\bf H} = H_x \hat {\bf x}$, but obviously
the result is independent of the field direction as long as it points along the surface plane, as we
have explicitly checked. In this case, the permittivity tensor of InSb adopts the form
\begin{equation}
\label{perm-tensor-Hx}
\hat \epsilon(H) = \left( \begin{array}{ccc} \epsilon_3 & 0 & 0 \\
0 & \epsilon_1(H) & -i\epsilon_2  \\ 0 & i\epsilon_2(H) & \epsilon_1(H) \end{array} \right) ,
\end{equation}
where the $\epsilon$'s are given by Eq.~(\ref{eq-epsilons}). Let us emphasize that in this case the transmission 
coefficient appearing in Eq.~(\ref{eq-net-Q}) depends both on the magnitude of the parallel wave and on its
direction, which makes the calculations more demanding. Let us also say that we consider here the same
parameter values for the $n$-doped InSb as in the example analyzed above.  

\begin{figure}[t]
\begin{center} \includegraphics[width=0.85\columnwidth,clip]{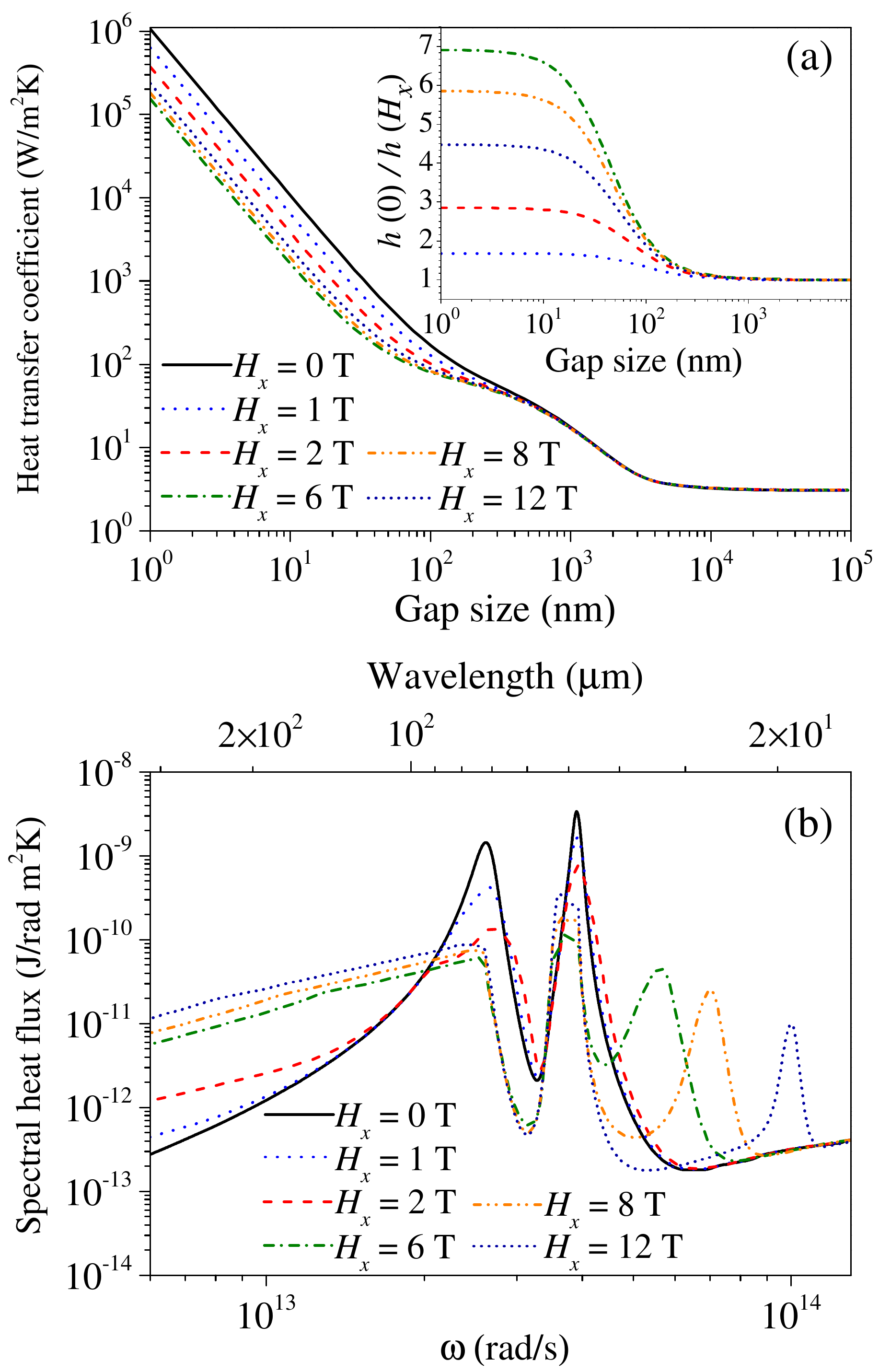} \end{center}
\caption{(a) Heat transfer coefficient for $n$-doped InSb as a function of the gap for different
values of the magnetic field applied along the surfaces of the plates. The inset shows
the ratio between the zero-field coefficient and the coefficient for different values of the
field in the near-field region. (b) The corresponding spectral heat flux as a function of
the frequency (and wavelength) for a gap of $d=10$ nm and different values of the parallel field.}
\label{fig-Hx}
\end{figure}
\begin{figure*}[t]
\includegraphics*[width=0.75\textwidth,clip]{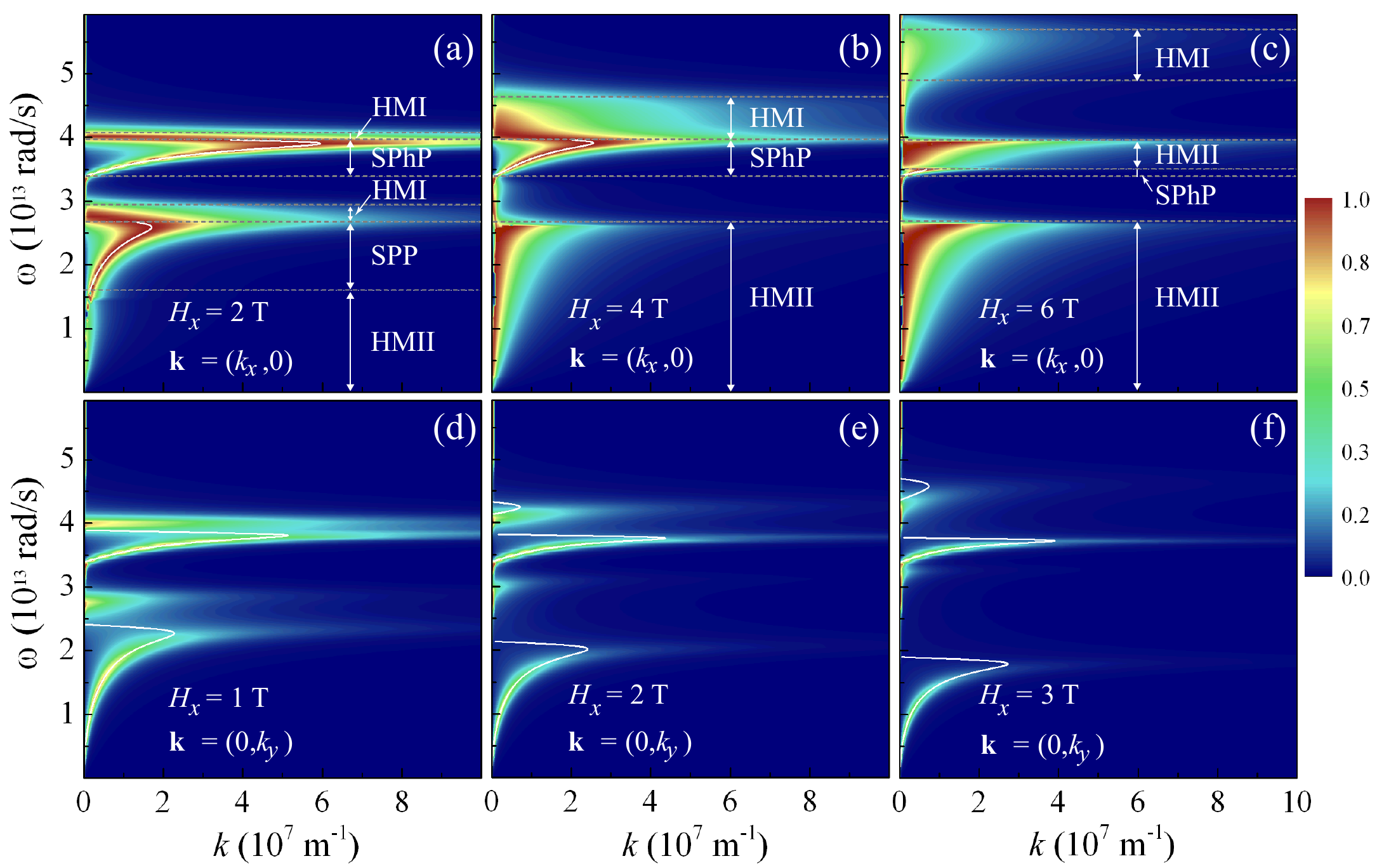}
\caption{The transmission coefficient for $p$-polarized waves as a function of the magnitude
of the parallel wave vector and frequency for InSb and a gap of $d=10$ nm. In all cases the field
is parallel to the plate surfaces, ${\bf H} = H_x \hat {\bf x}$. The panels (a-c) correspond to 
different values of the magnetic field for wave vectors parallel to field, ${\bf k} = (k_x,0)$, while 
panels (d-f) correspond to wave vectors perpendicular to the field, ${\bf k} = (0,k_y)$.
The horizontal dashed lines separate the regions where transmission is dominated by surface waves 
(SPPs and SPhPs) or hyperbolic modes of type I and II (HMI and HMII). The white solid lines correspond 
to the analytical dispersion relation of the surface waves of Eq.~(\ref{dr-Hz}) in panels (a-c) and
of Eq.~(\ref{dr-Hx}) in panels (d-f).}
\label{fig-trans-Hx}
\end{figure*}

The results for the magnetic field dependence of the heat transfer coefficient for the parallel
configuration are summarized in Fig.~\ref{fig-Hx}(a). As in the perpendicular case, the far-field
is barely affected by the magnetic field, the near-field thermal radiation is suppressed by the field, 
and at high fields the NFRHT tends to saturates. Interestingly, it saturates to the same value as in the
perpendicular configuration. In spite of the similarities, there are also important differences.
In this case, the NFRHT is much more sensitive to the field and a significant reduction is already 
achieved at 1 T. Notice also that in this case the heat transfer coefficient is clearly nonmonotonic
and the maximum reduction is reached at around 6 T. Finally, notice also that the reduction is more
pronounced than in the perpendicular case and the NFRHT can be diminished by up to a factor of 7 with
respect to the zero-field case, see inset of Fig.~\ref{fig-Hx}(a). This more pronounced 
reduction in the parallel configuration is also apparent in the spectral heat flux, as one can see 
in Fig.~\ref{fig-Hx}(b). Notice that also in this case there appears a high-frequency peak that is 
blue-shifted as the field increases. This peak appears at the cyclotron frequency and it has the 
same origin as in the perpendicular case.

Again, to understand this complex phenomenology, it is convenient to examine the transmission of
the $p$-polarized waves, which dominate the NFRHT for any field. Since in this case the transmission
also depends on the direction of {\bf k}, we choose to analyze the two most representative directions. In
the first one, the in-plane wave vector ${\bf k}$ is parallel to the field, \emph{i.e.}\ ${\bf k} = 
(k_x,0)$, and in the second one, ${\bf k}$ is perpendicular to the field,
\emph{i.e.}\ ${\bf k} = (0,k_y)$. The transmission of $p$-polarized waves
for these two directions is shown in Fig.~\ref{fig-trans-Hx} as a function of the magnitude of 
the wave vector and as a function of the frequency for different values of the field. As one can 
see, the transmission exhibits very different behaviors for these two directions. While for ${\bf k}
\parallel {\bf H}$ the situation resembles that of a perpendicular field (see discussion above), for
${\bf k} \perp {\bf H}$ it seems like the transmission is dominated by surface waves that are 
severely affected by the magnetic field (with the appearance of gaps in their dispersion relations).
These very different behaviors can be understood with an analysis of both the surface waves and the
propagating waves inside the material in these two situations. In the case ${\bf k} \parallel {\bf H}$,
one can show that a uniaxial approximation, similar to that discussed above, accurately reproduces 
the results for the transmission found in the exact calculation. In this case, the permittivity tensor 
can be approximated by $\hat \epsilon =$ \emph{diag}$[\epsilon_{xx},\epsilon_{zz},\epsilon_{zz}]$, where 
$\epsilon_{xx} = \epsilon_3$ and $\epsilon_{zz} = \epsilon_1$. Within this approximation, the dispersion 
relation of surface waves in the electrostatic limit $k \gg \omega/c$ is also given by Eq.~(\ref{dr-Hz}) 
(see Appendix D). As we show in Fig.~\ref{fig-trans-Hx}(a-c), this dispersion relation nicely describes
the structure of the transmission maxima in the regions where the surface waves can exist
($\epsilon_{xx},\epsilon_{zz} < 0$). On the other hand, as we show in Appendix C, the allowed 
values for the transverse component of the wave vector inside these uniaxial-like materials are 
given by $q_{\rm o} = \sqrt{\epsilon_{zz} \omega^2/c^2 - k^2}$ for ordinary waves and $q_{\rm e} 
= \sqrt{\epsilon_{xx} \omega^2/c^2 - k^2 \epsilon_{xx}/\epsilon_{zz}}$ for extraordinary waves.
Again, the dispersion of these extraordinary waves is of hyperbolic type when $\epsilon_{xx}$ and 
$\epsilon_{zz}$ have opposite signs. In Fig.~\ref{fig-trans-Hx}(a-c) we identify the frequency
regions where the HMs exist with the condition $\epsilon_{xx} \epsilon_{zz} < 0$, regions that
progressively dominate the transmission as the field increases. Thus, we see that for ${\bf k} 
\parallel {\bf H}$ the situation is very similar to that extensively discussed in the case
in which the field is perpendicular to the materials' surfaces.

On the contrary, the situation is very different for ${\bf k} \perp {\bf H}$. In this case, 
there are no HMs and no uniaxial approximation can describe the situation. As we show in 
Appendix C, the allowed $q$-values are given by $q_{\rm o,1} = \sqrt{\epsilon_{xx} \omega^2/c^2 - k^2}$ and
$q_{\rm o,2} = \sqrt{(\epsilon^2_{yy} + \epsilon^2_{yz}) \omega^2/(c^2 \epsilon_{yy}) - k^2}$, which 
both describe waves with no hyperbolic dispersion. On the other hand, the dispersion relation 
of the surface waves in the electrostatic limit is given by
\begin{equation}
\label{dr-Hx}
k_{\rm SW} = \frac{1}{2d} \ln \left( \frac{(\eta_{yy} -1+i\eta_{yz})(\eta_{yy} -1-i\eta_{yz})}
{(\eta_{yy} +1+i\eta_{yz})(\eta_{yy} +1-i\eta_{yz}) } \right) ,
\end{equation}
where $\eta_{yy} = \epsilon_{yy}/(\epsilon_{yy}^2+\epsilon_{yz}^2)$ and 
$\eta_{yz} = -\epsilon_{yz}/(\epsilon_{yy}^2+\epsilon_{yz}^2)$.
As we show in Fig.~\ref{fig-trans-Hx}(d-f), this dispersion relation explains the complex structure 
of the transmission maxima in this case. We emphasize that this dispersion relation is reciprocal in our
symmetric geometry and for this reason we only show results for $k_y > 0$. Notice that this dispersion is 
very sensitive to the magnetic field and already fields of the order of 1 T strongly affect the
surface waves. Notice also the appearance of gaps in the dispersion relations, a subject that has been
extensively discussed in the case of a single interface \cite{Kushwaha2001,Wallis1982}. Overall, the
field rapidly reduces the $k$-values of the surface waves and restricts the regions where they can exist.
This strong sensitivity of the surface waves with ${\bf k} \perp {\bf H}$ is the reason for the more 
pronounced reduction of the NFRHT for this field configuration. 

In general, for an arbitrary direction ${\bf k} = (k_x,k_y)$ the situation is somehow a combination of
the two types of behaviors just described. The complex interplay of these behaviors for different 
${\bf k}$-directions is responsible for the nonmonotonic dependence with magnetic field, 
along with the change in efficiency of the HMs upon varying the field. On the other
hand, at very high fields the cyclotron frequency becomes much larger than the plasma frequency and 
the off-diagonal elements of the permittivity tensor become negligible. At the same time, the field-dependent
terms in the diagonal elements also become very small. Thus, the systems effectively become uniaxial
and field-independent and the heat transfer is identical to the case in which the field is perpendicular.
Finally, in the far-field regime, the heat transfer is not sensitive to the magnetic field for the same
reason as in the perpendicular configuration.

Let us conclude this section with two brief comments. First, as it is obvious from the discussions above,
another way to modulate the NFRHT is by rotating the magnetic field, while keeping fixed its magnitude.
Actually, we find that for any field magnitude, the NFRHT is always smaller in the parallel configuration. 
Thus, one can increase or decrease the near-field thermal radiation by rotating appropriately the magnetic field.
Second, we have focused here in the case of doped InSb, but similar results can in principle be obtained
for other doped polar semiconductors such as GaAs, InAs, InP, PbTe, SiC, etc.

\section{Non-polar semiconductors: $\mbox{Si}$} \label{sec-IV}

In the previous section we have seen that when the field is parallel to the surfaces, one can have 
hyperbolic emitters, but the HMs always coexist to some degree with surface waves (even at the 
highest field). We show in this section that in the case of non-polar semiconductors, where
phonons do not play any role, it is possible to tune the system with a magnetic field to a situation
where only HMs contribute to the NFRHT. For this purpose, we choose Si as the material for the two
plates. As mentioned in the introduction, it has been predicted \cite{Fu2006,Rousseau2009b}, and 
experimentally tested \cite{Shi2013,Lim2015}, that in doped Si the NFRHT in the absence of field can be 
dominated by SPPs even at room temperature. Let us see now how this is modified upon applying a 
magnetic field.

\begin{figure}[t]
\begin{center} \includegraphics[width=0.85\columnwidth,clip]{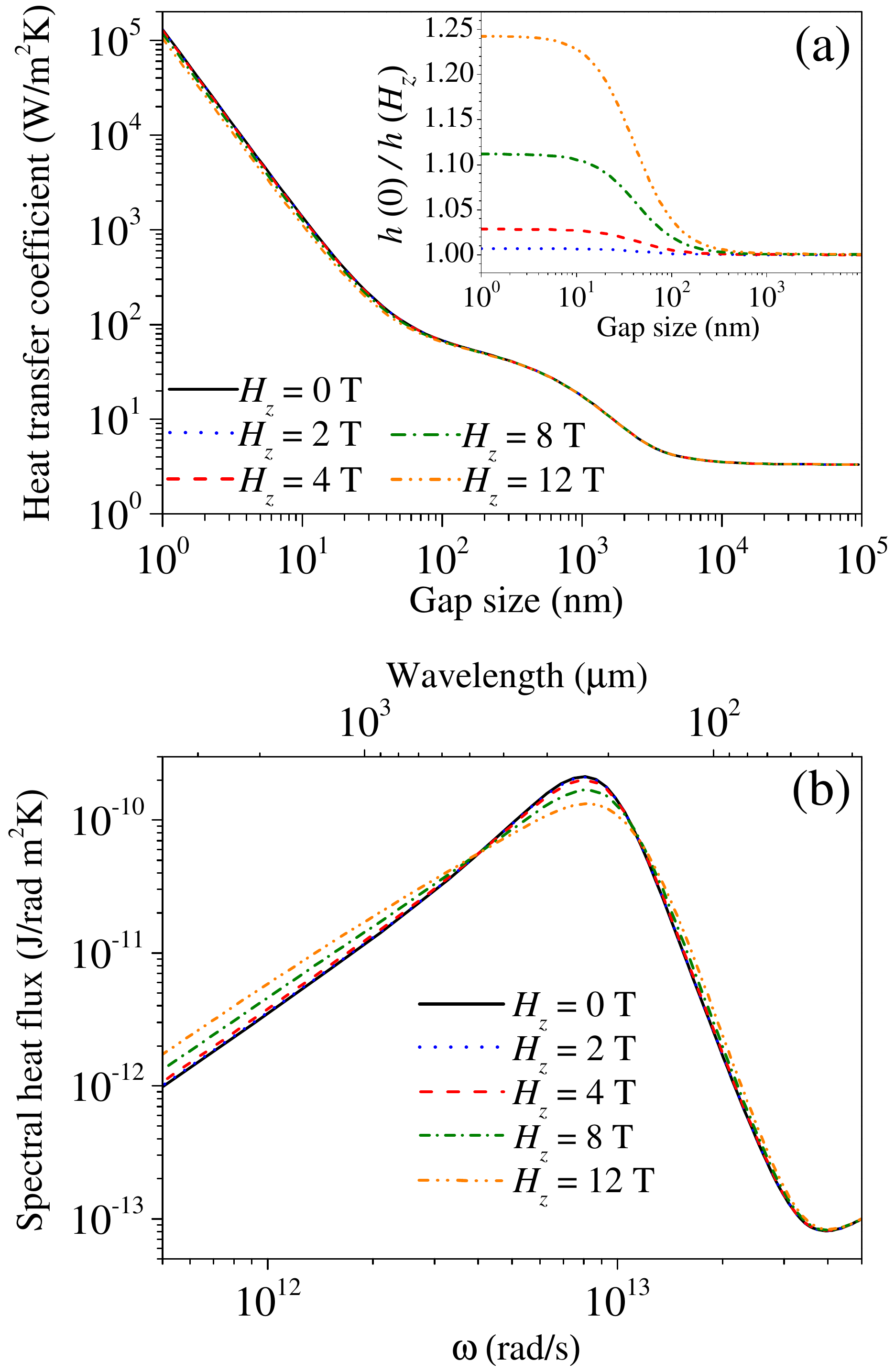} \end{center}
\caption{(a) Heat transfer coefficient for $n$-doped Si as a function of the gap for different
values of the magnetic field perpendicular to the plate surfaces. The inset shows
the ratio between the zero-field coefficient and the coefficient for different values of the
field in the near-field region. (b) The corresponding spectral heat flux as a function of
the frequency (and wavelength) for a gap of $d=10$ nm.}
\label{fig-Hz-Si}
\end{figure}
\begin{figure*}[t]
\includegraphics*[width=0.95\textwidth,clip]{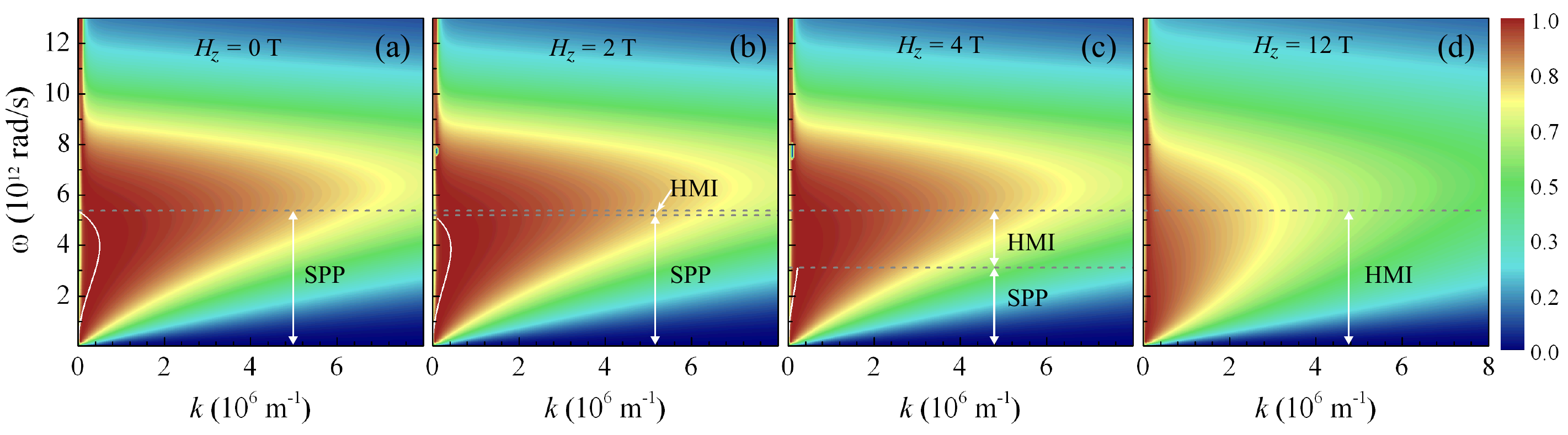}
\caption{The transmission coefficient for $p$-polarized waves as a function of the magnitude
of the parallel wave vector and frequency for Si and a gap of $d=10$ nm. The different panels
correspond to different values of the magnetic field that is perpendicular to the surfaces.
The horizontal dashed lines separate the regions where transmission is dominated by surface
plasmon polaritons (SPPs) or hyperbolic modes of type I (HMI). The white solid lines correspond
to the analytical SPP dispersion relation of Eq.~(\ref{dr-Hz}).}
\label{fig-trans-Hz-Si}
\end{figure*}
\begin{figure}[h!]
\begin{center} \includegraphics[width=0.85\columnwidth,clip]{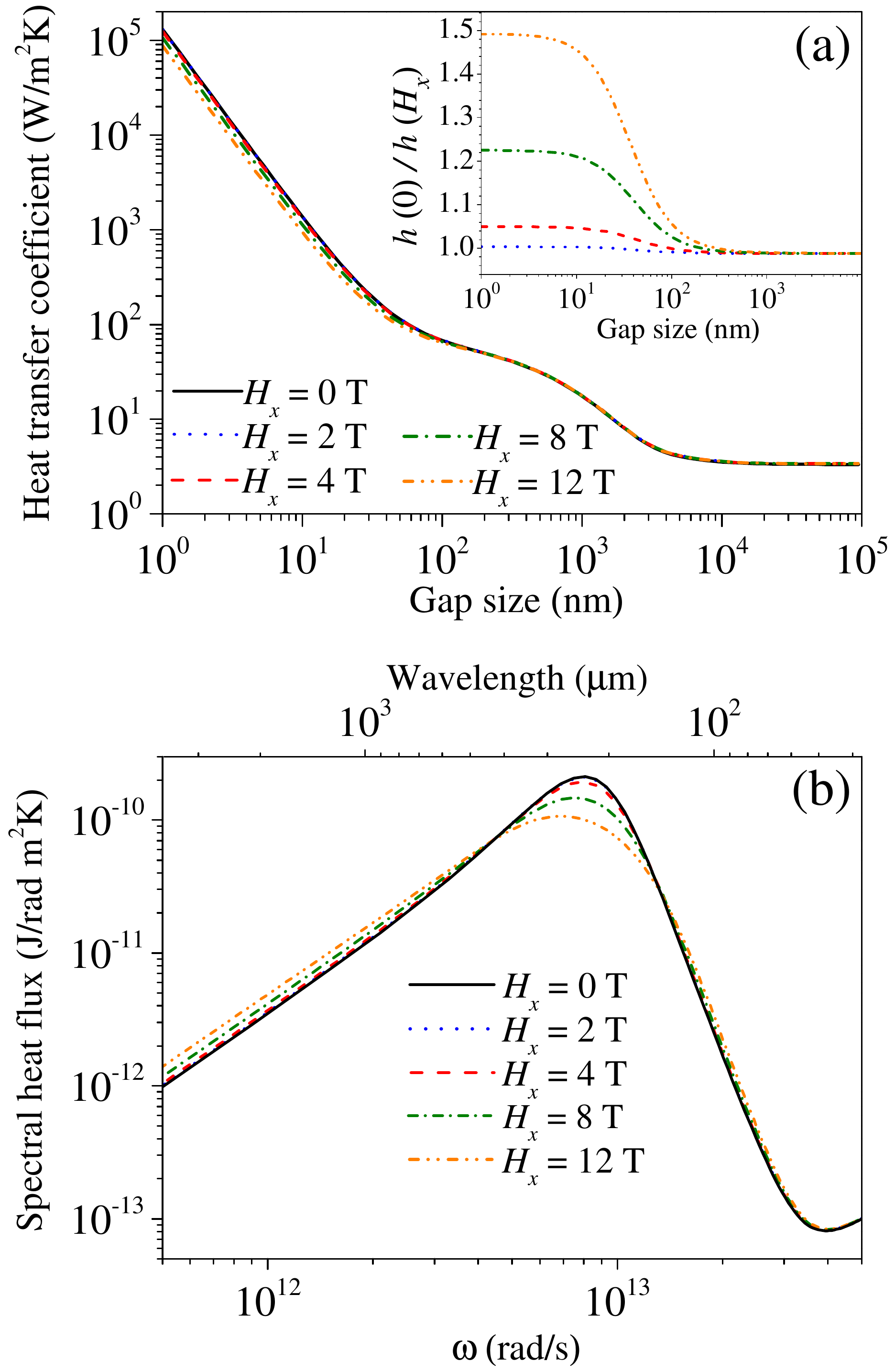} \end{center}
\caption{(a) Heat transfer coefficient for $n$-doped Si as a function of the gap for different
values of the magnetic field applied along the surfaces of the plates. The inset shows
the ratio between the zero-field coefficient and the coefficient for different values of the
field in the near-field region. (b) The corresponding spectral heat flux as a function of
the frequency (and wavelength) for a gap of $d=10$ nm.}
\label{fig-Hx-Si}
\end{figure}

The dielectric properties of doped Si are similar to those of InSb, the only difference being
the absence of a phonon contribution. Thus, the dielectric functions of Eq.~(\ref{eq-epsilons})
now read
\begin{eqnarray}
\epsilon_1(H) & = & \epsilon_{\infty} \left( 1 + \frac{\omega^2_p (\omega + i \gamma)}{\omega [\omega^2_c -
(\omega + i \gamma)^2]} \right) , \nonumber \\
\label{eq-epsilons-Si}
\epsilon_3 & = & \epsilon_{\infty} \left( 1 - \frac{\omega^2_p}{\omega (\omega + i \gamma)} \right) ,
\end{eqnarray}
while $\epsilon_2(H)$ remains unchanged. Using the results of Ref.~[\onlinecite{Fu2006}] for the
dielectric constant of doped Si, we focus on a room temperature case where the electron concentration 
is $n = 9.3 \times 10^{16}$ cm$^{-3}$, $\epsilon_{\infty} = 11.7$, $\gamma = 8.04 \times 10^{12}$ rad/s,
$m^{\ast}/m = 0.27$, and $\omega_p = 9.66 \times 10^{12}$ rad/s. We have chosen this doping level to have 
a situation in which the plasma frequency is not too high so that we can affect the NFRHT with a 
magnetic field, and not too low so that the NFRHT in the absence of field is still dominated by SPPs. 

The results for the heat transfer coefficient and spectral heat flux for a perpendicular magnetic
field are displayed in Fig.~\ref{fig-Hz-Si}. Although there are several features that are similar to
those of the InSb case, there are also some notable differences. To begin with, notice that now
higher fields are needed to see a significant reduction of the NFRHT (the required fields are around
an order of magnitude higher than for InSb) and the reduction factors are clearly more modest, 
see inset of Fig.~\ref{fig-Hz-Si}(a). This is mainly a consequence of the smaller cyclotron 
frequency in the Si case for a given field due to its larger effective mass. Another consequence
of the small cyclotron frequency is the fact that there is no sign of saturation of the NFRHT
for reasonable magnetic fields. On the other hand, the spectral heat flux at low fields is dominated 
this time by a single broad peak that originates from SPPs (see discussion below). As the field 
increases, the peak height is reduced and the peak itself is broadened and deformed. As we show in 
what follows, this behavior is due to the appearance of HMs that at high fields completely replace 
the surface waves.

Again, we can gain a further insight into these results by analyzing the transmission of the
$p$-polarized waves for different fields, which is illustrated in Fig.~\ref{fig-trans-Hz-Si}.
As one can see, the transmission is dominated by evanescent waves (in the vacuum gap) in a frequency 
region right below the plasma frequency. The origin of the structure of the transmission maxima
can be understood with the uniaxial approximation discussed above in the context of InSb. 
Again, this approximation reproduces very accurately all the results for arbitrary perpendicular 
fields (not shown here). Within this approximation, one can see that at low fields the transmission 
is dominated by SPPs, as we illustrate in Fig.~\ref{fig-trans-Hz-Si}(a-b) in which we have introduced 
the dispersion relation of the SPPs given by Eq.~(\ref{dr-Hz}). As soon as the magnetic magnetic field
becomes finite, the system starts to develop HMs of type I in a tiny frequency region right above the 
region of existence of the SPPs, see Fig.~\ref{fig-trans-Hz-Si}(b). The origin of these HMs is identical 
to that of the InSb case, but the main difference in this case is that upon increasing the field, one 
reaches a critical field value (of 4.36 T for this example) for which the surface waves cease to exist 
and the transmission is completely dominated by HMs turning the Si plates into ``pure" hyperbolic 
thermal emitters, see Fig.~\ref{fig-trans-Hz-Si}(c-d). 

For completeness, we have also studied the heat transfer in the parallel configuration and the
results for the heat transfer coefficient and spectral heat flux are shown in Fig.~\ref{fig-Hx-Si}.
In this case the results are rather similar to those of the perpendicular configuration. In particular,
contrary to the InSb case we do not find a nonmonotonic behavior. Moreover, the NFRHT reduction
is not much more pronounced than in the perpendicular case, although one can reach reduction factors
of 50\% for 12 T. Finally, saturation is not reached for these high fields for the same
reason as in the perpendicular configuration. As in the case of InSb, all these results can be
understood in terms of the modes that govern the near-field thermal radiation. In this sense, 
for a direction where ${\bf k} \parallel {\bf H}$, the SPPs that dominate the NFRHT at low fields
are progressively replaced by HMIs upon increasing the field and above 4.36 T they ``eat out" all
surface waves. On the contrary, for ${\bf k} \perp {\bf H}$ there are no HMs and the only magnetic 
field effect is the modification of the SPP dispersion relation. Again, the interplay between these 
two characteristic behaviors among the different ${\bf k}$-directions explains the evolution of the 
NFRHT with the field. 

Let us conclude this section by saying that the behavior reported here for Si could also
be observed for other non-polar semiconductors such as Ge.

\section{Outlook and conclusions} \label{sec-V}

The results reported in this work raise numerous interesting questions. Thus for instance, in all cases
analyzed so far, we have found that the magnetic field reduces the NFRHT as compared to the zero-field
result. Is there any fundamental argument that forbids a magnetic-field-induced enhancement? In principle,
there is no such an argument. The reduction that we have found in doped semiconductors is due to the
fact that the we have explored cases where surface waves, which are extremely efficient, dominate the 
NFRHT in the absence of field. In this sense, one may wonder if a field-induced enhancement could 
take place in a situation where the NFRHT in the absence of field is dominated by standard frustrated
internal reflection modes, as it happens in metals \cite{Chapuis2008}. Obviously, metals are out of
the question due to their huge plasma frequency, but one can investigate non-polar semiconductors with
a low doping level. Indeed, we have done it for the case of Si and again, we find that the magnetic field
reduces the NFRHT and moreover, exceedingly high fields are required to see any significant effect. Of
course, we have by no means exhausted all possibilities and, for instance, we have not explored 
asymmetric situations with different materials. Thus, the question remains of whether the application
of a magnetic field can under certain circumstances enhance the near-field thermal radiation.

The discovery in this work of the induction of hyperbolic modes upon the application of a magnetic field
may also have important consequences for layered structures involving thin films. Recently, it has been
demonstrated that thin films made of polar dielectrics may support NFRHT enhancements comparable to those
of bulk samples due to the excitation of SPhPs \cite{Song2015b}. Since hyperbolic modes have a propagating
character inside the material, they may be severely affected in a thin film geometry by the presence of
a substrate. Thus, one could expect much more dramatic magnetic-field effects in systems coated with 
semiconductor thin films.  

Obviously, the question remains of whether one can modulate the NFRHT with a magnetic field in other classes
of materials. For instance, since a magneto-optical activity is required, what about ferromagnetic materials?
Ideally, one could imagine to tune the NFRHT by playing around with the relative orientation of the 
magnetization, following the spin-valve experiments in the context of spintronics. 

Another question of general interest for the field of metamaterials is if a doped semiconductor under a
magnetic field could exhibit the plethora of exotic optical properties reported in hybrid hyperbolic 
metamaterials \cite{Poddubny2013,Shekhar2014}. We have shown here that it can behave as a hyperbolic thermal 
emitter, but can it also exhibit negative refraction or be used do to subwavelength imaging and focusing 
in the infrared? These are very important questions that we are currently pursuing.

So in summary, we have presented in this work a very detailed theoretical analysis of the influence
of a magnetic field in the NFRHT. By considering the simple case of two parallel plates, we have
demonstrated that for doped semiconductors the near-field thermal radiation can be strongly modified 
by the application of an external magnetic field. In particular, we have shown that the magnetic field 
may significantly reduce the NFRHT and the reduction in polar semiconductors can be as large as 700\%
at room temperature. Moreover, we have shown that when the field is perpendicular to the parallel plates, 
doped semiconductors become ideal hyperbolic thermal emitters with highly tunable properties. This 
provides a unique opportunity to explore the physics of thermal radiation in this class of metamaterials 
without the need to resort to complex hybrid structures. Finally, all the predictions of this work are 
amenable to measurements with the present experimental techniques, and we are convinced that the multiple 
open questions that this work raises will motivate many new theoretical and experimental studies of this subject.

\begin{acknowledgments}

This work was financially supported by the Colombian agency COLCIENCIAS, the Spanish Ministry 
of Economy and Competitiveness (Contract No.\ FIS2014-53488-P), and the Comunidad de Madrid (Contract 
No.\ S2013/MIT-2740). V.F-H. acknowledges financial support from ``la Caixa" Foundation and 
F.J.G-V. from the European Research Council (ERC-2011-AdG proposal no. 290981).

\end{acknowledgments}

\appendix

\section{Scattering matrix approach for anisotropic multilayer systems}

Our analysis of the radiative heat transfer in the presence of a static magnetic field is based 
on the combination of Rytov's fluctuational electrodynamics (FE) and a scattering matrix formalism
that describes the propagation of electromagnetic waves in multilayer systems made of optically 
anisotropic materials. As we show in Appendix B, the radiative heat transfer can be expressed in 
terms of the scattering matrix of our system. Thus, it is convenient to first discuss in this 
appendix the scattering matrix approach employed in this work ignoring for the moment the fluctuating 
currents that generate the thermal radiation. Later in Appendix B, we show how this approach can be 
combined with FE. We follow here Ref.~[\onlinecite{Caballero2012}], which presents a generalization 
of the formalism introduced by Whittaker and Culshaw in Ref.~[\onlinecite{Whittaker1999}] for 
isotropic systems.

Let us first describe the Maxwell's equations to be solved. Assuming a harmonic time dependence 
$\exp(-i\omega t)$, the Maxwell's equations for non-magnetic materials and in the absence of 
currents adopt the following form: ${\bf \nabla} \cdot \epsilon_0 \hat \epsilon {\bf E} = 0$, 
${\bf \nabla} \cdot {\bf H} = 0$, ${\bf \nabla} \times {\bf H} = -i \omega \epsilon_0 \hat 
\epsilon {\bf E}$, and ${\bf \nabla} \times {\bf E} =  i \omega \mu_0 {\bf H}$, where the permittivity
is in general a tensor given by Eq.~(\ref{perm-tensor}). The first Maxwell's equation is automatically 
satisfied if the third one is fulfilled, and the second one can be satisfied by expanding the magnetic 
field in terms of basis functions with zero divergence. Following Ref.~[\onlinecite{Whittaker1999}], 
it is convenient to introduce the rescaling: $\omega \epsilon_0 {\bf E} \to {\bf E}$ and $\sqrt{\mu_0 \epsilon_0} 
\omega = \omega/c \to \omega$. Thus, the final two equations to be solved are
\begin{eqnarray}
\label{Meq1}
{\bf \nabla} \times {\bf H} & = & -i \hat \epsilon {\bf E} , \\
\label{Meq2}
{\bf \nabla} \times {\bf E} & = & i \omega^2 {\bf H} .
\end{eqnarray}

We consider here a planar multilayer system grown along the $z$-direction in which 
the tensor $\hat \epsilon$ is constant inside every layer, \emph{i.e.}\ it is independent
of the in-plane coordinates ${\bf r} \equiv (x,y)$. Thus, for an in-plane wave vector 
${\bf k} \equiv (k_x, k_y)$, we can write the fields as
\begin{equation}
{\bf H} ({\bf r}, z) = \boldsymbol{h}(z) e^{i{\bf k} \cdot {\bf r}} \;\;\mbox{and}\;\;
{\bf E} ({\bf r}, z) = \boldsymbol{e}(z) e^{i{\bf k} \cdot {\bf r}} .
\end{equation}
With this notation, Eqs.~(\ref{Meq1}) and (\ref{Meq2}) can be rewritten as
\begin{eqnarray}
\label{Amp1}
i k_y h_z(z) - h^{\prime}_y(z) & = & -i \sum_j \epsilon_{xj} e_j(z) \\
\label{Amp2}
h^{\prime}_x(z) - i k_x h_z(z) & = & -i \sum_j \epsilon_{yj} e_j(z) \\
\label{Amp3}
i k_x h_y(z) - i k_y h_x(z) & = & -i \sum_j \epsilon_{zj} e_j(z) ,
\end{eqnarray}
and
\begin{eqnarray}
\label{Farad1}
i k_y e_z(z) - e^{\prime}_y(z) & = & i \omega^2 h_x(z) \\
\label{Farad2}
e^{\prime}_x(z) - i k_x e_z(z) & = & i \omega^2 h_y(z) \\
\label{Farad3}
i k_x e_y(z) - i k_y e_x(z) & = & i \omega^2 h_z(z) ,
\end{eqnarray}
where the primes stand for $\partial_z$.

Now our task is to solve the Maxwell's equations for an unbounded layer. 
For this purpose, we write the magnetic field $\boldsymbol{h}(z)$ as follows
\begin{equation}
\boldsymbol{h}(z) = e^{iqz} \left\{ \phi_x {\bf \hat x} + \phi_y {\bf \hat y} -
\frac{1}{q} (k_x \phi_x + k_y \phi_y ) {\bf \hat z} \right\} ,
\end{equation}
where ${\bf \hat x}$, ${\bf \hat y}$, and ${\bf \hat z}$ are the Cartesian unit
vectors and $q$ is the $z$-component of the wave vector. Here, $\phi_x$
and $\phi_y$ are the expansion coefficients to be determined by
substituting into Maxwell's equations. Notice that this expression satisfies
${\bf \nabla} \cdot {\bf H} = 0$. Now, it is convenient to rewrite the previous
expression in the vector notation:
\begin{equation}
\boldsymbol{h}(z) = e^{iqz} \left( \phi_x, \phi_y , - \frac{1}{q}
(k_x \phi_x + k_y \phi_y ) \right)^T . \label{h-kspace}
\end{equation}
With this notation, Eqs.~(\ref{Amp1}-\ref{Amp3}) can be written as
\begin{equation}
\label{Amp-matrix}
\hat {\cal C} \boldsymbol{h}(z) = \hat \epsilon \boldsymbol{e}(z) ,\;\;\mbox{where}\;\;
\hat {\cal C} = \left( \begin{array}{ccc}
0 & q & - k_y \\
-q & 0 &  k_x \\
k_y & - k_x & 0 \end{array} \right).
\end{equation}
On the other hand, Eqs.~(\ref{Farad1}-\ref{Farad3}) adopt now the form
\begin{equation}
\hat {\cal C}^{T} \boldsymbol{e}(z) = \omega^2 \boldsymbol{h}(z) .
\label{Farad-matrix}
\end{equation}
From Eq.~(\ref{Amp-matrix}) we obtain the following expression for the electric
field
\begin{equation}
\label{e-kspace}
\boldsymbol{e}(z) = \hat \eta \hat {\cal C} \boldsymbol{h}(z) ,
\end{equation}
where $\hat \eta = \hat \epsilon^{-1}$. Substituting this expression in
Eq.~(\ref{Farad-matrix}) we obtain the following equation for the magnetic field
\begin{equation}
\hat {\cal C}^{T} \hat \eta \hat {\cal C} \boldsymbol{h}(z) =
\omega^2 \boldsymbol{h}(z) ,
\end{equation}
which defines an eigenvalue problem for $\omega^2$. Indeed, only two of the three
identities obtained from this equation, one for each ${\bf \hat x}$, ${\bf \hat y}$,
and ${\bf \hat z}$, are independent. From the first two identities, and using
Eq.~(\ref{h-kspace}), we obtain the following equations determining the allowed
values for $q$
\begin{equation}
\left( \hat {\cal A}_2 q^2 + \hat {\cal A}_1 q + \hat {\cal A}_0 + \hat {\cal A}_{-1} \frac{1}q
\right) \phi = 0 ,
\label{rational-eigen}
\end{equation}
where $\phi = (\phi_x, \phi_y)^{T}$ and the $2\times 2$ matrices $\hat {\cal A}_n$
are defined by
\begin{widetext}
\begin{eqnarray}
& & \hat {\cal A}_2 = \left( \begin{array}{cc} \eta_{yy} & -\eta_{yx} \\
- \eta_{xy} & \eta_{xx} \end{array} \right) , \;\;
\hat {\cal A}_1 = \hat {\cal A}^{(a)}_1 + \hat {\cal A}^{(b)}_1 =
\left( \begin{array}{cc} - k_y \eta_{zy} & k_y \eta_{zx} \\
k_x \eta_{zy} & - k_x \eta_{zx} \end{array} \right) +
\left( \begin{array}{cc} - k_y \eta_{yz} & k_x \eta_{yz} \\
k_y \eta_{xz} & - k_x \eta_{xz} \end{array} \right) , \nonumber \\
& & \hat {\cal A}_0 = \hat {\cal A}^{(a)}_0 + \hat {\cal A}^{(b)}_0 -\omega^2 \hat 1 =
\left( \begin{array}{cc} k^2_y \eta_{zz} & - k_x k_y \eta_{zz} \\
- k_x k_y \eta_{zz} & k^2_x \eta_{zz} \end{array} \right) +
\left( \begin{array}{cc} k^2_x \eta_{yy} - k_x k_y \eta_{yx} &
k_x k_y \eta_{yy} - k^2_y \eta_{yx} \\
k_x k_y \eta_{xx} - k^2_x \eta_{xy}  &
k^2_y \eta_{xx} - k_x k_y \eta_{xy} \end{array}
\right) - \omega^2 \left( \begin{array}{cc} 1 & 0 \\ 0 & 1  \end{array} \right) ,
\nonumber \\
& & \hat {\cal A}_{-1} = \left( \begin{array}{cc} k^2_y k_x \eta_{zx}
- k^2_x k_y \eta_{zy} & k^3_y \eta_{zx}
- k^2_y k_x \eta_{zy} \\ k^3_x \eta_{zy}
- k^2_x k_y \eta_{zx} & k^2_x k_y \eta_{zy}
- k^2_y k_x \eta_{zx} \end{array} \right) . \label{A-def}
\end{eqnarray}
This eigenvalue problem leads to the following quartic secular equation:
$\sum^4_{n=0} D_n q^n = 0$, where the coefficients are given by
\begin{eqnarray}
D_4 & = & \eta_{xx} \eta_{yy} - \eta_{xy} \eta_{yx} ,
\nonumber \\
D_3 & = & k_x \left[ \eta_{xy} \eta_{yz} + \eta_{yx} \eta_{zy} - \eta_{yy}
(\eta_{xz} + \eta_{zx}) \right] + k_y \left[ \eta_{yx} \eta_{xz} + \eta_{xy} \eta_{zx} - 
\eta_{xx} (\eta_{yz} + \eta_{zy}) \right] ,
\nonumber \\
D_2 & = & k^2_x \left[ \eta_{yy} (\eta_{xx} + \eta_{zz}) - \eta_{xy} \eta_{yx}
- \eta_{yz} \eta_{zy} \right] + k^2_y \left[ \eta_{xx} (\eta_{yy} + \eta_{zz}) -
\eta_{xy} \eta_{yx} - \eta_{xz} \eta_{zx} \right] \nonumber \\ & & + k_x k_y \left[
\eta_{xz} (\eta_{yz} + \eta_{zy}) + \eta_{yz} (\eta_{zx} - \eta_{xz})
\eta_{zz} (\eta_{xy} + \eta_{yx}) \right] - \omega^2 (\eta_{xx} + \eta_{yy}) ,
\nonumber \\
D_1 & = & k^3_x \left[ \eta_{xy} \eta_{yz} + \eta_{yx} \eta_{zy} - \eta_{yy}
(\eta_{xz} + \eta_{zx}) \right] + k^3_y \left[ \eta_{yx} \eta_{xz} +
\eta_{xy} \eta_{zx} - \eta_{xx} (\eta_{yz} + \eta_{zy}) \right] \nonumber \\ & &
+ k^2_x k_y \left[ \eta_{xy} \eta_{zx} + \eta_{xz} \eta_{yx} - \eta_{xx}
(\eta_{yz} + \eta_{zy}) \right] + k^2_y k_x
\left[ \eta_{yx} \eta_{zy} + \eta_{yz} \eta_{xy} - \eta_{yy} (\eta_{xz} + \eta_{zx})
\right] \nonumber \\ & & + \omega^2 \left[ k^2_x(\eta_{xz} + \eta_{zx}) + k^2_y
(\eta_{yz} + \eta_{zy}) \right] ,
\nonumber \\
D_0 & = & k^4_x (\eta_{yy} \eta_{zz} - \eta_{yz} \eta_{zy}) + k^4_y (\eta_{xx}
\eta_{zz} - \eta_{xz} \eta_{zx})
+ k^3_x k_y \left[ \eta_{xz} \eta_{zy} + \eta_{yz} \eta_{zx} - \eta_{zz}
(\eta_{xy} + \eta_{yx}) \right] \nonumber \\ & &
+ k^3_y k_x \left[ \eta_{yz} \eta_{zx} + \eta_{xz} \eta_{zy} - \eta_{zz}
(\eta_{yx} + \eta_{xy}) \right]
+ k^2_x k^2_y \left[ \eta_{zz} (\eta_{xx} + \eta_{yy}) + \eta_{xy} \eta_{yx} -
\eta_{xz} \eta_{zx} - \eta_{yz} \eta_{zy} \right] \nonumber \\ & &
+ \omega^2 \left[ \omega^2 - k^2_x ( \eta_{yy} + \eta_{zz}) - k^2_y (\eta_{xx} + \eta_{zz})
+ k_x k_y (\eta_{xy} + \eta_{yx}) \right] .
\end{eqnarray}
\end{widetext}
In general, this secular equation has to be solved numerically, but in many situations
of interest the allowed values for $q$ can be obtained analytically (see Appendix C).
The solution of Eq.~(\ref{rational-eigen}) provides four complex eigenvalues for $q$,
two lie in the upper half of the complex plane and the other two in the lower half.

The next step toward the solution of the Maxwell's equations in a multilayer structure is 
the determination of the fields in the different layers. This can be done by expressing 
the fields as a combination of forward and backward propagating waves with wave numbers 
$q_n$ (with $n=1,2$), and complex amplitudes $a_n$ and $b_n$, respectively. These amplitudes
will be determined later by using the boundary conditions at the interfaces and surfaces
of the multilayer structure. Since the boundary conditions are simply the continuity of
the in-plane field components, we focus here on the analysis of the field components $e_x$,
$e_y$, $h_x$, and $h_y$. From Eq.~(\ref{h-kspace}), the in-plane components of $\boldsymbol{h}$
can be expanded in terms of propagating waves as follows
\begin{eqnarray}
\left( \begin{array}{c} h_x(z) \\ h_y(z) \end{array} \right) & = & \sum^2_{n=1} \left\{
\left( \begin{array}{c} \phi_{x_n} \\ \phi_{y_n} \end{array} \right) e^{iq_nz} a_n
\right. \nonumber \\ & & \hspace*{5mm} \left. + \left( \begin{array}{c} \varphi_{x_n} \\
\varphi_{y_n} \end{array} \right) e^{-ip_n(d-z)} b_n \right\} ,
\end{eqnarray}
where $d$ is the thickness of the layer. Here, $a_n$ is the coefficient of the forward
going wave at the $z=0$ interface, and $b_n$ is the backward going wave at $z=d$.
On the other hand, $q_n$ correspond to the eigenvalues of Eq.~(\ref{rational-eigen})
with $\mbox{Im} \{q_n\} > 0$ and $p_n$ are the eigenvalues with $\mbox{Im} \{p_n\} < 0$. 

To simplify the notation, we now define two $2 \times 2$ matrices
$\hat \Phi_+$ and $\hat \Phi_-$ whose columns are the vectors $\phi_n$ and $\varphi_n$,
respectively. Moreover, we define the diagonal $2 \times 2$ matrices
$\hat {\rm f}_+(z)$ and $\hat {\rm f}_-(d-z)$, such that $[\hat {\rm f}_+(z)]_{nn}
= e^{iq_nz}$ and $[\hat {\rm f}_-(d-z)]_{nn} = e^{-ip_n(d-z)}$, and the $2$-dimensional
vectors $\boldsymbol{h}_{\parallel}(z) = (h_x(z), h_y(z))^{T}$, $\boldsymbol{a} = (a_1,a_2)^{T}$, 
and $\boldsymbol{b} = (b_1,b_2)^{T}$. In terms of these quantities, the in-plane
magnetic-field components become
\begin{equation}
\boldsymbol{h}_{\parallel}(z) = \hat \Phi_+ \hat {\rm f}_+(z) \boldsymbol{a} + 
\hat \Phi_- \hat {\rm f}_-(d-z) \boldsymbol{b} .
\label{hp-exp}
\end{equation}

Similarly, from Eq.~(\ref{e-kspace}) it is straightforward to show that the in-plane 
components of the electric field, $\boldsymbol{e}_{\parallel}(z) = (-e_y(z), e_x(z))^{T}$, are given by
\begin{eqnarray}
\boldsymbol{e}_{\parallel}(z) & = & \left( \hat {\cal A}^{(b)}_0 \hat \Phi_+ \hat q^{-1} +
\hat {\cal A}^{(b)}_1 \hat \Phi_+ + \hat {\cal A}_2 \hat \Phi_+ \hat q \right) 
\hat {\rm f}_+(z) \boldsymbol{a} \nonumber \\ & & \hspace*{-1.5cm}
+ \left( \hat {\cal A}^{(b)}_0 \hat \Phi_- \hat p^{-1} + \hat {\cal A}^{(b)}_1 \hat \Phi_-
+ \hat {\cal A}_2 \hat \Phi_- \hat p \right) \hat {\rm f}_-(d-z) \boldsymbol{b} ,
\label{ep-exp}
\end{eqnarray}
where the $\hat {\cal A}$'s are defined in Eq.~(\ref{A-def}) and we have defined the
$2\times 2$ diagonal matrices $\hat q$ and $\hat p$ such that $\hat q_{nn} =
q_n$ and $\hat p_{nn} = p_n$.

We can now combine Eq.~(\ref{hp-exp}) and (\ref{ep-exp}) into a single expression
as follows
\begin{eqnarray}
\label{M-def}
\left( \begin{array}{c} \boldsymbol{e}_{\parallel}(z) \\ \boldsymbol{h}_{\parallel}(z) \end{array} \right) & = & \hat M \left(
\begin{array}{c} \hat {\rm f}_+(z) \boldsymbol{a} \\ \hat {\rm f}_-(d-z) \boldsymbol{b} \end{array} \right) \\
& = & \left( \begin{array}{cc} \hat M_{11} & \hat M_{12} \\ \hat M_{21} & \hat M_{22} \end{array} \right)
\left( \begin{array}{c} \hat {\rm f}_+(z) \boldsymbol{a} \\ \hat {\rm f}_-(d-z) \boldsymbol{b} \end{array} \right)
, \nonumber
\end{eqnarray}
where the $2 \times 2$ matrices $M_{ij}$ are defined as
\begin{eqnarray}
\hat M_{11} & = & \hat {\cal A}^{(b)}_0 \hat \Phi_+ \hat q^{-1} + \hat {\cal A}^{(b)}_1 \hat \Phi_+
+ \hat {\cal A}_2 \hat \Phi_+ \hat q , \nonumber \\
\hat M_{12} & = & \hat {\cal A}^{(b)}_0 \hat \Phi_- \hat p^{-1} + \hat {\cal A}^{(b)}_1 \hat \Phi_-
+ \hat {\cal A}_2 \hat \Phi_- \hat p , \nonumber \\
\hat M_{21} & = & \hat \Phi_+ , \;\;\; \hat M_{22} =  \hat \Phi_- .
\end{eqnarray}

The final step in our calculation is to use the scattering matrix ($S$-matrix) to compute 
the field amplitudes needed to describe the different relevant physical quantities. By 
definition, the $S$-matrix relates the vectors of the amplitudes of forward and backward 
going waves, $\boldsymbol{a}_l$ and $\boldsymbol{b}_l$, where $l$ now denotes the layer, in the
different layers of the structure as follows
\begin{equation}
\left( \begin{array}{c} \boldsymbol{a}_l \\ \boldsymbol{b}_{l^{\prime}} \end{array} \right) = 
\hat S(l^{\prime},l) \left( \begin{array}{c} \boldsymbol{a}_{l^{\prime}} \\ \boldsymbol{b}_l \end{array} \right) =
\left( \begin{array}{cc} \hat S_{11} & \hat S_{12} \\ \hat S_{21} & \hat S_{22} \end{array} \right)
\left( \begin{array}{c} \boldsymbol{a}_{l^{\prime}} \\ \boldsymbol{b}_l \end{array} \right) .
\label{S-def}
\end{equation}

The field amplitudes in two consecutive layers are related via the continuity of the in-plane 
components of the fields in every interface and surface. If we consider the interface
between the layer $l$ and the layer $l+1$, this continuity leads to
\begin{equation}
\left( \begin{array}{c} \boldsymbol{e}_{\parallel}(d_l) \\ \boldsymbol{h}_{\parallel}(d_l) \end{array} \right)_l =
\left( \begin{array}{c} \boldsymbol{e}_{\parallel}(0) \\ \boldsymbol{h}_{\parallel}(0) \end{array} \right)_{l+1} ,
\end{equation}
where $d_l$ is the thickness of layer $l$. From this condition, together with Eq.~(\ref{M-def}), 
it is easy to show that the amplitudes in layers $l$ and $l+1$ are related by the interface matrix 
$\hat I(l,l+1) = \hat M^{-1}_l \hat M_{l+1}$ in the following way
\begin{eqnarray}
\left( \begin{array}{c} \hat f^+_{l} \boldsymbol{a}_l \\ \boldsymbol{b}_l \end{array} \right) 
& = & \hat I(l,l+1) \left( \begin{array}{c} \boldsymbol{a}_{l+1} \\ \hat f^-_{l+1} \boldsymbol{b}_{l+1} 
\end{array} \right) \nonumber \\ & = &
\left( \begin{array}{cc} \hat I_{11} & \hat I_{12} \\ \hat I_{21} & \hat I_{22} \end{array} \right)
\left( \begin{array}{c} \boldsymbol{a}_{l+1} \\ \hat f^-_{l+1} \boldsymbol{b}_{l+1} \end{array} \right) ,
\label{eq-I-matrix}
\end{eqnarray}
where $\hat f^+_{l} = \hat {\rm f}_{l,+}(d_l)$ and $\hat f^-_{l+1} = \hat
{\rm f}_{l+1,-}(d_{l+1})$.

Now, with the help of the interface matrices, the $S$-matrix can be calculated
in an iterative way as follows. The matrix $\hat S(l^{\prime},l+1)$ can be calculated
from $\hat S(l^{\prime},l)$ using the definition of $\hat S(l^{\prime},l)$ in Eq.~(\ref{S-def})
and the interface matrix $\hat I(l,l+1)$. Eliminating $\boldsymbol{a}_l$ and $\boldsymbol{b}_l$ 
we obtain the relation between $\boldsymbol{a}_{l^{\prime}}$, $\boldsymbol{b}_{l^{\prime}}$ and 
$\boldsymbol{a}_{l+1}$, $\boldsymbol{b}_{l+1}$, from which $\hat S(l^{\prime},l+1)$ can be 
constructed. This reasoning leads to the following iterative relations
\begin{eqnarray}
\hat S_{11}(l^{\prime},l+1) & = & \left[\hat I_{11} - \hat f^+_{l} \hat S_{12}(l^{\prime},l) \hat I_{21}
\right]^{-1} \hat f^+_{l} \hat S_{11}(l^{\prime},l) \nonumber \\
\hat S_{12}(l^{\prime},l+1) & = & \left[\hat I_{11} - \hat f^+_{l} \hat S_{12}(l^{\prime},l) \hat I_{21}
\right]^{-1} \nonumber \\ & & \times \left( \hat f^+_{l} \hat S_{12}(l^{\prime},l) \hat I_{22}
- \hat I_{12} \right) \hat f^-_{l+1} \nonumber \\
\hat S_{21}(l^{\prime},l+1) & = & \hat S_{22}(l^{\prime},l) \hat I_{21} \hat S_{11}(l^{\prime},l+1) +
\hat S_{21}(l^{\prime},l) \nonumber \\
\hat S_{22}(l^{\prime},l+1) & = & \hat S_{22}(l^{\prime},l) \hat I_{21} \hat S_{12}(l^{\prime},l+1) +
\nonumber \\ & & \hat S_{22}(l^{\prime},l) \hat I_{22} \hat f^-_{l+1} .
\label{eq-iterative}
\end{eqnarray}
Starting from $\hat S(l^{\prime},l^{\prime}) = 1$, one can apply the previous recursive relations to 
a layer at a time to build up $\hat S(l^{\prime},l)$. Let us conclude this appendix by saying that 
from the knowledge of the $S$-matrix one can easily compute the field amplitudes in every layer and,
in turn, the fields everywhere in the system \cite{Whittaker1999}.

\section{Thermal radiation in anisotropic multilayer systems}

In this appendix we show how the scattering matrix approach of Appendix A can be used
to describe the thermal radiation between planar multilayer systems made of anisotropic materials.
For this purpose, we first discuss how a generic emission problem can be formulated in the 
framework of the $S$-matrix formalism and then, we show how such a formulation can be used 
to describe the thermal emission of a multilayer system. 

\subsection{Emission in the scattering matrix approach}

For concreteness, let us assume that there is a set of oscillating point sources, with 
harmonic time dependence, occupying the whole plane defined by $z=z^{\prime}$. The corresponding
electric current density ${\bf J}$ is given by
\begin{equation}
{\bf J}({\bf r},z) = {\bf J}_0 \delta(z-z^{\prime}) = \boldsymbol{j}_0 e^{i {\bf k} \cdot 
{\bf r}} \delta(z-z^{\prime}),
\end{equation}
where $\boldsymbol{j}_0({\bf k}) = {\bf J}_0 e^{-i {\bf k} \cdot {\bf r}}$. This current density
enters as a source term in Amp\`ere's law, Eq.~(\ref{Meq1}), which now becomes ${\bf \nabla} 
\times {\bf H} = {\bf J} -i \hat \epsilon {\bf E}$, while Eq.~(\ref{Meq2}) (Faraday's law) 
remains unchanged. Thus, Eqs.~(\ref{Amp1}-\ref{Amp3}) adopt now the following form 
\begin{eqnarray}
\label{Ampj1}
i k_y h_z(z) - h^{\prime}_y(z) & = & j_{0x} \delta(z-z^{\prime}) -i \sum_j \epsilon_{xj} e_j(z) \\
\label{Ampj2}
h^{\prime}_x(z) - i k_x h_z(z) & = & j_{0y} \delta(z-z^{\prime}) -i \sum_j \epsilon_{yj} e_j(z) \\
\label{Ampj3}
i k_x h_y(z) - i k_y h_x(z) & = & j_{0z} \delta(z-z^{\prime}) -i \sum_j \epsilon_{zj} e_j(z) . \hspace{0.7cm}
\end{eqnarray}

The presence of the source term induces discontinuities in the fields across the plane $z=z^{\prime}$,
as we proceed to show. First, let us consider the effect of the in-plane components of the 
current density by putting $j_z = 0$. To cancel the singular term due to the source in Eqs.~(\ref{Ampj1})
and (\ref{Ampj2}), there must be discontinuities in $h_x$ and $h_y$ at $z=z^{\prime}$ equal to 
$j_{0y}$ and $-j_{0x}$, respectively. All the other field components are continuous, except for
$e_z$ that exhibits a discontinuity equal to $(k_x j_{0x} + k_y j_{0y})/\epsilon_{zz}$ in virtue 
of Eq.~(\ref{Ampj3}). Let us analyze now the role of the perpendicular component of ${\bf J}$ 
by putting $j_{0x} = j_{0y} = 0$. From Eq.~(\ref{Ampj3}), it is clear that in this case $e_z$
must contain a singularity to cancel the singular term associated to the current source, that
is $e_z(z) = -i(j_{0z}/\epsilon_{zz}) \delta(z-z^{\prime}) +$ non-singular parts. This introduces
singular terms in the left-hand side of the Maxwell Eqs.~(\ref{Farad1}) and (\ref{Farad2}),
which are cancelled by discontinuities in $e_x$ and $e_y$ equal to $k_x j_{0z}/\epsilon_{zz}$
and $k_y j_{0z}/\epsilon_{zz}$, respectively. Additionally, it is obvious from Eqs.~(\ref{Ampj1})
and (\ref{Ampj2}) that $h_x$ and $h_y$ acquired discontinuities equal to $-\epsilon_{yz} j_{0z}/
\epsilon_{zz}$ and $\epsilon_{xz} j_{0z}/\epsilon_{zz}$, respectively. Defining the following
vectors
\begin{eqnarray}
\boldsymbol{p}_{\parallel} & = & (j_{0y} - \epsilon_{yz} j_{0z}/ \epsilon_{zz}, 
-j_{0x} + \epsilon_{xz} j_{0z}/ \epsilon_{zz})^T \\
\boldsymbol{p}_{z} & = & (- k_y j_{0z}/ \epsilon_{zz}, k_x j_{0z}/ \epsilon_{zz})^T, 
\end{eqnarray}
the boundary conditions on the in-plane components of the fields are thus
\begin{eqnarray}
\boldsymbol{e}_{\parallel}(z^{\prime +}) - \boldsymbol{e}_{\parallel}(z^{\prime -}) & = & \boldsymbol{p}_z \nonumber \\
\boldsymbol{h}_{\parallel}(z^{\prime +}) - \boldsymbol{h}_{\parallel}(z^{\prime -}) & = & \boldsymbol{p}_{\parallel} . 
\end{eqnarray}

These discontinuity conditions can now be combined with the $S$-matrix formalism of the previous
appendix to calculate the emission throughout the system. Let us consider that the emission plane
defines the interface between layers $l$ and $l+1$ in our multilayer structure. Thus, the boundary
conditions in this interface become
\begin{equation}
\left( \begin{array}{c} \boldsymbol{e}_{\parallel}(0) \\ \boldsymbol{h}_{\parallel}(0) \end{array} \right)_{l+1} -
\left( \begin{array}{c} \boldsymbol{e}_{\parallel}(d_l) \\ \boldsymbol{h}_{\parallel}(d_l) \end{array} \right)_{l} =
\left( \begin{array}{c} \boldsymbol{p}_{z} \\ \boldsymbol{p}_{\parallel} \end{array} \right) .
\end{equation}
Using now the expression of the fields in terms of the layer matrices ($\hat M$'s), see Eq.~(\ref{M-def}),
we can write
\begin{equation}
\label{bcem}
\hat M_{l+1} \left( \begin{array}{c} \boldsymbol{a}_{l+1} \\ \hat f^-_{l+1} \boldsymbol{b}_{l+1} \end{array} \right) -
\hat M_l \left( \begin{array}{c} \hat f^+_{l} \boldsymbol{a}_l \\ \boldsymbol{b}_l \end{array} \right) = 
\left( \begin{array}{c} \boldsymbol{p}_{z} \\ \boldsymbol{p}_{\parallel} \end{array} \right) .
\end{equation}

The external boundary conditions for an emission problem is that there should be only outgoing waves, 
that is $\boldsymbol{a}_0 = \boldsymbol{b}_N = \boldsymbol{0}$, where 0 denotes here the first layer 
of the structure and $N$ the last one. Using the definitions of the $S$-matrices $\hat S(0,l)$ and 
$\hat S(l+1,N)$ from Eq.~(\ref{S-def}), it follows that
\begin{eqnarray}
\label{ampr1}
\boldsymbol{a}_l & = & \hat S_{12}(0,l) \boldsymbol{b}_l \\
\label{ampr4}
\boldsymbol{b}_{l+1} & = & \hat S_{21}(l+1,N) \boldsymbol{a}_{l+1} .
\end{eqnarray}
Substituting for $\boldsymbol{a}_l$ and $\boldsymbol{b}_{l+1}$ from Eqs.~(\ref{ampr1}) and (\ref{ampr4})
in Eq.~(\ref{bcem}) and rearranging things, we arrive at the following central result
\begin{widetext}
\begin{equation}
\label{eq-emiss}
\left( \begin{array}{cc} \hat M_{11,l+1} + \hat M_{12,l+1} \hat f^-_{l+1} \hat S_{21}(l+1,N) &
- [ \hat M_{12,l} + \hat M_{11,l} \hat f^+_{l} \hat S_{12}(0,l) ] \\
\hat M_{21,l+1} + \hat M_{22,l+1} \hat f^-_{l+1} \hat S_{21}(l+1,N) &
- [ \hat M_{22,l} + \hat M_{21,l} \hat f^+_{l} \hat S_{12}(0,l) ] \end{array} \right)
\left( \begin{array}{c} \boldsymbol{a}_{l+1} \\ \boldsymbol{b}_l \end{array} \right) =
\left( \begin{array}{c} \boldsymbol{p}_{z} \\ \boldsymbol{p}_{\parallel} \end{array} \right) ,
\end{equation}
\end{widetext}
which allows us to compute the field amplitudes on the left and on the right-hand side of the
emitting plane. From the solution of this matrix equation we can compute the field amplitude
everywhere inside and outside the multilayer structure from the knowledge of the scattering
matrix. 

\subsection{Radiative heat transfer}

Let us now show that the previous results can be used to describe the radiative heat transfer. 
First of all, we need to specify the properties of the electric currents that generate the thermal 
radiation. In the framework of fluctuational electrodynamics \cite{Rytov1989}, the thermal emission 
is generated by random currents ${\bf J}$ inside the material. While the statistical average of 
these currents vanishes, \emph{i.e.}\ $\langle {\bf J} \rangle = 0$, their correlations are given 
by the fluctuation-dissipation theorem \cite{Landau1980,Keldysh1994}
\begin{eqnarray}
\label{eq-FDT}
\langle J_k({\bf R},\omega) J^{\ast}_l({\bf R}^{\prime},\omega^{\prime}) \rangle & = & 
\frac{4\epsilon_0 \omega c}{\pi} \Theta(\omega,T) \delta({\bf R}-{\bf R}^{\prime}) \delta(\omega - \omega^{\prime}) \times
\nonumber \\ & & \left[ \epsilon_{kl}({\bf R},\omega) - \epsilon^{\ast}_{lk}({\bf R},\omega) \right]/(2i) , \hspace{1cm} 
\end{eqnarray}
where ${\bf R} = ({\bf r},z)$ and $\Theta(\omega,T) = \hbar \omega c/ [\exp(\hbar \omega c / k_{\rm B}T) -1]$, 
$T$ being the absolute temperature. Let us remind the reader that with the rescaling introduced at the 
beginning of Appendix A, $\omega$ has dimensions of wave vector in our notation. Notice that in the 
expression of $\Theta(\omega,T)$ a term equal to $\hbar \omega c/2$ that accounts for vacuum fluctuations
has been omitted since it does not affect the neat radiation heat flux. Notice also that 
we are using here the most general form of this theorem that is suitable for non-reciprocal systems. 
The fact that the current correlations are local in space and diagonal in frequency space reduces the problem 
of the thermal radiation to the description of the emission by point sources for a given frequency, parallel 
wave vector, and position inside the structure. Thus, we can directly apply the results derived in the previous 
subsection.

\begin{figure}[t]
\begin{center} \includegraphics[width=0.9\columnwidth,clip]{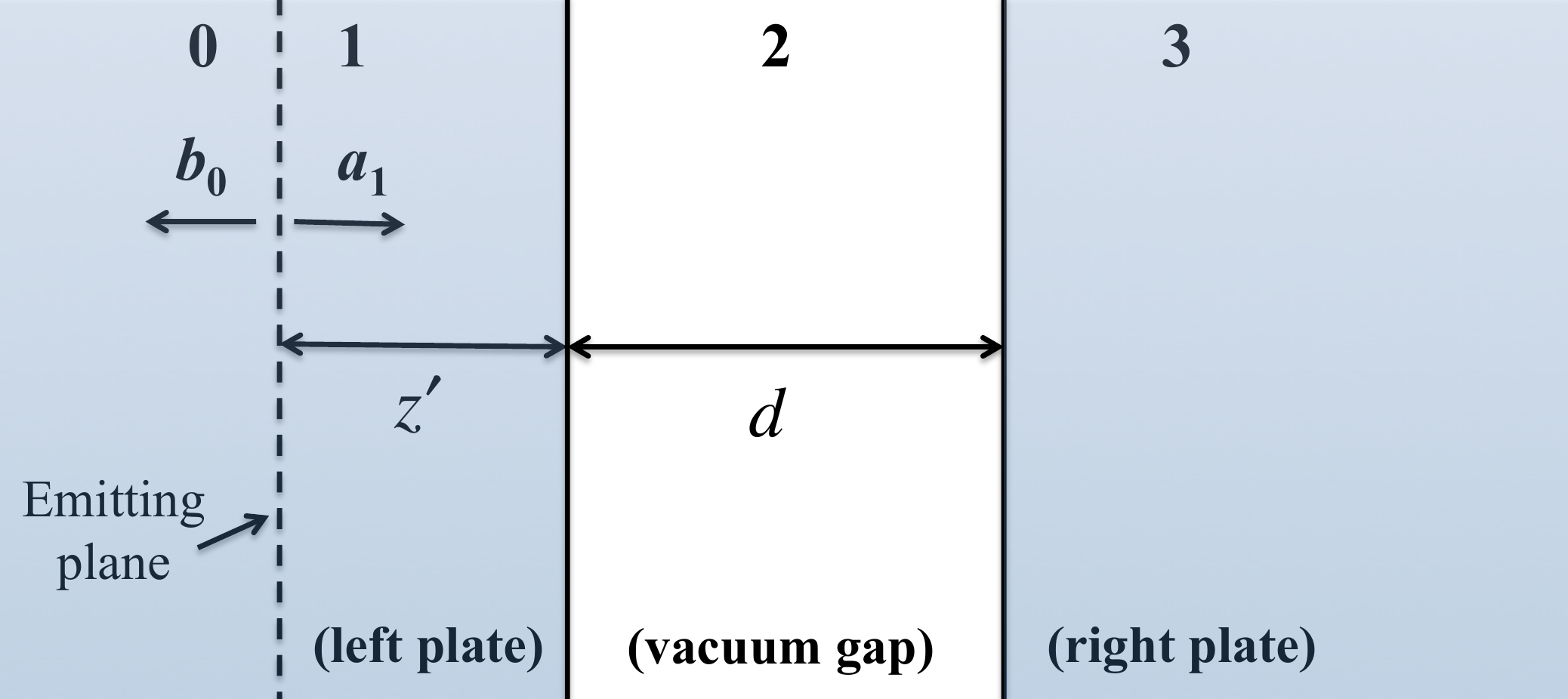} \end{center}
\caption{Two parallel plates separated by a vacuum gap of width $d$. The vertical dashed line inside 
the left plate indicates the position of an emitting plane that contains the radiation sources that 
generate the field amplitudes $\boldsymbol{b}_0$ and $\boldsymbol{a}_1$.} 
\label{setup}
\end{figure}

Let us now consider our system of study, namely two parallel plates at temperatures $T_1$ and $T_3$ separated 
by a vacuum gap of width $d$, see Fig.~\ref{setup}. Our strategy to compute the net radiative heat transfer 
between the two plates follows closely that of the seminal work by Polder and Van Hove \cite{Polder1971}.
First, we compute the radiation power per unit of area transferred from the left plate to the 
right one, $Q_{1 \rightarrow 3}$. For this purpose, we first compute the statistical average of the
$z$-component of the Poynting vector describing the power emitted from a plane located at $z=z^{\prime}$
inside the left plate for a given frequency and parallel wave vector and then, we integrate integrate 
the result over all possible values of $z^{\prime}$, $\omega$, and ${\bf k}$, \emph{i.e.}\
\begin{equation}
\label{eq-Q13}
Q_{1 \rightarrow 3}(d,T_1) = \int^{\infty}_0 d\omega \int d{\bf k} \int^{\infty}_0 dz^{\prime} 
\langle S_z(\omega, {\bf k}, z^{\prime}) \rangle .
\end{equation}
A similar calculation for the power $Q_{3 \rightarrow 1}$ transferred from the right plate to the 
left one completes the computation of the net transferred power per unit of area.

Let us focus now on the analysis of the power emitted by a plane inside the left plate, see Fig.~\ref{setup}. 
This emitting plane defines a fictitious interface between layers 0 and 1, which are both inside the left
plate. To determine the power emitted to the right plate we first compute the field amplitudes 
$\boldsymbol{a}_1$ on the right hand side of the plane. For this purpose we make of use of 
Eq.~(\ref{eq-emiss}), where in this case $l=0$ and $N=3$. Taking into account that $\hat S_{12}(0,0)
= 0$, it is straightforward to show that 
\begin{eqnarray}
\boldsymbol{a}_1 & = & [\hat M_{11,1} - \hat M_{12,1} \hat M^{-1}_{22,1} \hat M_{21,1}]^{-1} \boldsymbol{p}_z
+ \nonumber \\ & & [\hat M_{21,1} - \hat M_{22,1} \hat M^{-1}_{12,1} \hat M_{11,1}]^{-1} 
\boldsymbol{p}_{\parallel} \nonumber \\ & = & 
[\hat M^{-1}_1]_{11} \boldsymbol{p}_z + [\hat M^{-1}_1]_{12} \boldsymbol{p}_{\parallel} .
\label{eq-amp-a1}
\end{eqnarray}

To compute $Q_{1 \to 3}$, it is convenient to calculate the Poynting vector in the vacuum gap. For
this purpose, we need the field amplitudes in that layer. From Eq.~(\ref{S-def}), it is easy to
deduce that these amplitudes are given in terms of $\boldsymbol{a}_1$ as follows
\begin{equation}
\boldsymbol{a}_2 = \hat D \hat S_{11}(1,2) \boldsymbol{a}_1 ,
\label{eq-field-amp2a}
\end{equation}
where $\hat D \equiv [\hat 1 - \hat S_{12}(1,2) \hat S_{21}(2,3)]^{-1}$ and
\begin{eqnarray}
\boldsymbol{b}_2 & = & \left[\hat 1 - \hat S_{21}(2,3) \hat S_{12}(1,2) \right]^{-1} 
\hat S_{21}(2,3) \hat S_{11}(1,2) \boldsymbol{a}_1 \nonumber \\ 
& = & \hat S_{21}(2,3) \boldsymbol{a}_2 .
\label{eq-field-amp2b}
\end{eqnarray}
It is worth stressing that the different elements of the scattering matrix that appear in
the previous expressions can be factorized into scattering matrices $\tilde S$ containing only 
information about the interfaces of the layered system, which are basically the Fresnel 
coefficients of the structure, and phase factors describing the propagation between these 
interfaces. In particular, from Eq.~(\ref{eq-iterative}) it is easy to show that
\begin{eqnarray}
\hat S_{11}(1,2) & = & \tilde S_{11}(1,2) \hat f^+_1(z^{\prime}) \\
\hat S_{12}(1,2) & = & \tilde S_{12}(1,2) e^{iq_2 d} \\
\hat S_{21}(2,3) & = & \tilde S_{21}(2,3) e^{iq_2 d} ,
\end{eqnarray}
where $q_2 = \sqrt{\omega^2 - k^2}$ is the $z$-component of the wave vector in the vacuum gap and
\begin{equation}
\hat f^+_1(z^{\prime}) = \left( \begin{array}{cc} e^{iq_{1,1} z^{\prime}} & 0 \\
0 & e^{iq_{2,1} z^{\prime}} \end{array} \right) .
\end{equation}
Here, $q_{i,1}$ (with $i=1,2$) are the $z$-components of the two allowed wave vectors in the medium 1.
On the other hand, the $\tilde S$-matrices can be computed directly from the interface matrices as follows
[see Eq.~(\ref{eq-iterative})]
\begin{eqnarray}
\tilde S_{11}(1,2) & = & \hat I^{-1}_{11}(1,2) \\
\tilde S_{12}(1,2) & = & -\hat I^{-1}_{11}(1,2) \hat I_{12}(1,2) \\
\tilde S_{21}(2,3) & = & \hat I_{21}(2,3) \hat I^{-1}_{11}(2,3) .
\end{eqnarray}

In terms of the amplitudes $\boldsymbol{a}_2$ and $\boldsymbol{b}_2$, the fields in the vacuum
gap at $z=0$ are given by [see Eq.~(\ref{M-def})]
\begin{equation}
\left( \begin{array}{c} \boldsymbol{e}_{\parallel}(0) \\ \boldsymbol{h}_{\parallel}(0) \end{array} \right)_2 =
\left( \begin{array}{c} \hat M_{11,2} \left[ \boldsymbol{a}_2 - e^{i q_2 d} \boldsymbol{b}_2 \right] \\
\boldsymbol{a}_2 + e^{i q_2 d} \boldsymbol{b}_2 \end{array} \right) ,
\end{equation}
where we have used that $\hat M_{12,2} = - \hat M_{11,2}$, valid for any isotropic system. Thus, the
$z$-component of the Poynting vector evaluated at $z=0$ in the vacuum gap reads
\begin{eqnarray}
S_z(\omega,{\bf k}, z^{\prime}) & = & \frac{1}{4\omega} \sqrt{\frac{\mu_0}{\epsilon_0}} \left\{
\boldsymbol{h}^{\dagger}_{\parallel}(0) \boldsymbol{e}_{\parallel}(0) +
\boldsymbol{e}^{\dagger}_{\parallel}(0) \boldsymbol{h}_{\parallel}(0) \right\}_2 \nonumber \\ 
& = & \frac{1}{4\omega} \sqrt{\frac{\mu_0}{\epsilon_0}} \left\{ \boldsymbol{a}^{\dagger}_2 
\left(\hat M_{11,2} + \hat M^{\dagger}_{11,2} \right) \boldsymbol{a}_2 - \right. \nonumber \\ & &
e^{i(q_2 - q^{\ast}_2)d} \boldsymbol{b}^{\dagger}_2 \left( \hat M_{11,2} + \hat M^{\dagger}_{11,2} 
\right) \boldsymbol{b}_2 + \nonumber \\ & &
e^{-iq^{\ast}_2 d} \boldsymbol{b}^{\dagger}_2 \left( \hat M_{11,2} - \hat M^{\dagger}_{11,2} \right) 
\boldsymbol{a}_2 - \nonumber \\ & & \left. e^{iq_2 d} \boldsymbol{a}^{\dagger}_2 \left( \hat M_{11,2} - 
\hat M^{\dagger}_{11,2} \right) \boldsymbol{b}_2 \right\} .
\label{eq-PV}
\end{eqnarray}
Moreover, since
\begin{equation}
\hat M_{11,2} = \frac{1}{q_2} \left( \begin{array}{cc} \omega^2 - k^2_y & k_x k_y \\
k_x k_y & \omega^2 - k^2_x \end{array} \right) \equiv \frac{1}{q_2} \hat A 
\end{equation}
and $q_2$ is either real (for $k < \omega$) or purely imaginary (for $k > \omega$), Eq.~(\ref{eq-PV})
reduces to
\begin{eqnarray}
\label{eq-PV2}
S_z(\omega,{\bf k}, z^{\prime}) = \frac{1}{2 q_2 \omega} \sqrt{\frac{\mu_0}{\epsilon_0}} \times \hspace{4cm} & & \\ 
\left\{ \begin{array}{ll} \boldsymbol{a}^{\dagger}_2 \hat A \boldsymbol{a}_2 - \boldsymbol{b}^{\dagger}_2 
\hat A \boldsymbol{b}_2, & k < \omega \\ e^{-iq^{\ast}_2 d} \boldsymbol{b}^{\dagger}_2 \hat A \boldsymbol{a}_2 - 
e^{iq_2 d} \boldsymbol{a}^{\dagger}_2 \hat A \boldsymbol{b}_2, & k > \omega , \end{array} \right. & & \nonumber
\end{eqnarray}
where the first term provides the contribution of propagating waves and the second one corresponds to
the contribution of evanescent waves.

From this point on, the rest of the calculation is pure algebra and we will not describe it here in 
detail. Let us simply say that the basic idea is to use Eqs.~(\ref{eq-field-amp2a}) and (\ref{eq-field-amp2b}) 
to express the Poynting vector in Eq.~(\ref{eq-PV2}) in terms of the field amplitude $\boldsymbol{a}_1$. 
Then, using Eq.~(\ref{eq-amp-a1}) and the fluctuation-dissipation theorem of Eq.~(\ref{eq-FDT}) one can calculate 
the statistical average of the Poynting vector. Let us mention that the calculation can be greatly simplified 
by rotating every $2 \times 2$ matrix appearing in the problem from the Cartesian basis ($x$-$y$) to the basis of 
$s$- and $p$-polarized waves. This can be done via the unitary matrix
\begin{equation}
\hat R \equiv \frac{1}{k} \left( \begin{array}{cc} k_x & k_y \\ k_y & -k_x \end{array} \right) ,
\end{equation}
which is the matrix that defines the transformation that diagonalizes the matrix $\hat A$, \emph{i.e.}\
\begin{equation}
\hat A_d \equiv \hat R \hat A \hat R = \left( \begin{array}{cc}
\omega^2 & 0 \\ 0 & q^2_2 \end{array} \right) .
\end{equation}
Finally, after integrating over all possible values of $\omega$, ${\bf k}$, and $z^{\prime}$, see 
Eq.~(\ref{eq-Q13}), one arrives at the following result for the power per unit of area transferred 
from the left plate to the right one
\begin{equation}
Q_{1 \rightarrow 3}(d,T_1) = \int^{\infty}_{0} \frac{d \omega}{2\pi} \Theta(\omega,T_1)
\int \frac{d{\bf k}}{(2\pi)^2} \tau(\omega,{\bf k},d) ,
\end{equation}
where $\tau(\omega,{\bf k},d)$ is the total transmission coefficient of the electromagnetic modes and
it is given by 
\begin{widetext}
\begin{equation}
\label{eq-trans-app}
\tau(\omega,{\bf k},d) = \left\{ \begin{array}{ll}
\mbox{Tr} \left\{ [\hat 1 - \bar S_{12}(1,2) \bar S^{\dagger}_{12}(1,2) ] \bar D^{\dagger}
[\hat 1 - \bar S^{\dagger}_{21}(2,3) \bar S_{21}(2,3) ] \bar D \right\}, & k < \omega
\;\; \mbox{(propagating waves)} \\
\mbox{Tr} \left\{ [\bar S_{12}(1,2) - \bar S^{\dagger}_{12}(1,2) ] \bar D^{\dagger}
[\bar S^{\dagger}_{21}(2,3) - \bar S_{21}(2,3) ] \bar D \right\} e^{-2|q_2|d}, & k > \omega
\;\; \mbox{(evanescent waves)} \end{array} \right. .
\end{equation}
\end{widetext}
Here, the $2 \times 2$ matrices indicated by a bar are defined as follows
\begin{eqnarray}
\bar D & \equiv & \hat A^{1/2}_d \hat R \hat D \hat R \hat A^{-1/2}_d \\
\bar D^{\dagger} & \equiv & \hat A^{-1/2}_d \hat R \hat D^{\dagger} \hat R \hat A^{1/2}_d .
\end{eqnarray}

Following a similar reasoning, one can compute the power per unit of area transfer from the 
right plate to the left one and the final result reads
\begin{equation}
Q_{3 \rightarrow 1}(d,T_3) = \int^{\infty}_{0} \frac{d \omega}{2\pi} \Theta(\omega,T_3)
\int \frac{d{\bf k}}{(2\pi)^2} \tau(\omega,{\bf k},d) ,
\end{equation}
where $\tau(\omega,{\bf k},d)$ is also given by Eq.~(\ref{eq-trans-app}). Thus, the net power
per unit of area exchanged by the plates is given by Eqs.~(\ref{eq-net-Q}) and (\ref{eq-trans-man}) 
in section \ref{sec-II}. To conclude, let us stress that in the manuscript $\omega$ is meant
to be an angular frequency.

\section{Dispersion relations}

In this appendix we provide the solution of the eigenvalue problem of Eqs.~(\ref{rational-eigen}) 
and (\ref{A-def}) that give the dispersion relations of the electromagnetic modes that can exist 
inside the materials considered in this work. In particular, we focus here on three cases of special
interest for our discussions in the main body of the manuscript. 

\emph{Case 1:} $\hat {\epsilon} = diag[\epsilon_{xx}, \epsilon_{xx}, \epsilon_{zz}]$. This 
situation is of relevance for the case in which the magnetic field is perpendicular to the plate 
surfaces, see section \ref{sec-III}A. In this case, the allowed $q$-values are given by
\begin{equation}
\label{eq-qs-case1}
q^2_{\rm o} = \epsilon_{xx} \omega^2 - k^2, \;\; 
q^2_{\rm e} = \epsilon_{xx} \omega^2 - k^2 \epsilon_{xx}/\epsilon_{zz} .
\end{equation}

\emph{Case 2:} $\hat {\epsilon} = diag[\epsilon_{xx}, \epsilon_{zz}, \epsilon_{zz}]$. This 
situation is relevant for the case in which the magnetic field is parallel to the plate surfaces,
see section \ref{sec-III}B, and the allowed $q$-values are given by
\begin{equation}
\label{eq-qs-case2}
q^2_{\rm o} = \epsilon_{zz} \omega^2 - k^2,  \;\;
q^2_{\rm e} = \epsilon_{xx} \omega^2 - k^2 \epsilon_{xx}/\epsilon_{zz} .
\end{equation}

\emph{Case 3:} the diagonal elements of $\hat {\epsilon}$ are $\epsilon_{xx}$ and $\epsilon_{yy} =
\epsilon_{zz}$, while the only non-vanishing off-diagonal elements are $\epsilon_{yz} = - \epsilon_{zy}$. 
This situation is relevant for the case in which the magnetic field is parallel to the plate surfaces,
see section \ref{sec-III}B. In this case, the allowed $q$-values adopt the following form
\begin{equation}
\label{eq-qs-case3}
q^2_{\rm o,1} = \epsilon_{xx} \omega^2 - k^2,  \;\;
q^2_{\rm o,2} = (\epsilon^2_{yy} + \epsilon^2_{yz}) \omega^2/\epsilon_{yy} - k^2 .
\end{equation}
%


\section{Surface electromagnetic modes}

We briefly describe here how we determine the dispersion relation of the surface electromagnetic 
modes in our system and we also provide the results for some configurations of special interest.

Let us consider a structure containing $N$ planar layers. From Eq.~(\ref{eq-I-matrix}), 
it is easy to show that the field amplitudes in layers $l$ and $l+1$ are related as follows
\begin{eqnarray}
\left( \begin{array}{c} \boldsymbol{a}_l \\ \boldsymbol{b}_l \end{array} \right)
& = & \left( \begin{array}{cc} \hat f^+_{l} & \hat 0 \\ \hat 0 & \hat 1 \end{array} \right)^{-1} 
\hat I(l,l+1) \left( \begin{array}{cc} \hat 1 & \hat 0 \\ \hat 0 & \hat f^-_{l+1} \end{array} \right)
\left( \begin{array}{c} \boldsymbol{a}_{l+1} \\ \boldsymbol{b}_{l+1} \end{array} \right) \nonumber \\
& \equiv & \hat I^{\prime}(l,l+1) \left( \begin{array}{c} \boldsymbol{a}_{l+1} \\ \boldsymbol{b}_{l+1} 
\end{array} \right) .
\end{eqnarray}
Now, using this relation recursively we can relate the field amplitudes in the first and last layers
as follows
\begin{equation}
\left( \begin{array}{c} \boldsymbol{a}_1 \\ \boldsymbol{b}_1 \end{array} \right)
= \left[ \prod^{N-1}_{l=1} \hat I^{\prime}(l,l+1) \right]
\left( \begin{array}{c} \boldsymbol{a}_{N} \\ \boldsymbol{b}_{N} \end{array} \right) 
\equiv \hat I^{S} \left( \begin{array}{c} \boldsymbol{a}_{N} \\ \boldsymbol{b}_{N} 
\end{array} \right) .
\end{equation}
The condition for an eigenmode of the system is that $\boldsymbol{a}_1 = \boldsymbol{b}_{N} = 
{\bf 0}$, which from the previous equation implies that $\hat I^{S}_{11} \boldsymbol{a}_{N} = {\bf 0}$.
The condition for having a non-trivial solution of this equation is that $\det \hat I^{S}_{11} = 0$,
which is the condition that surface electromagnetic modes must satisfy. In our plate-plate geometry, 
the $4 \times 4$ matrix $\hat I^{S}$ is simply given by
\begin{equation}
\hat I^{S} = \hat I(1,2) \left( \begin{array}{cc} e^{-iq_2d} \hat 1 & \hat 0 \\
\hat 0 & e^{iq_2d} \hat 1 \end{array} \right) \hat I(2,3) ,
\end{equation}
where let us recall that $q_2 = \sqrt{\omega^2 - k^2}$. Thus, the condition for an
eigenmode of the system reads
\begin{equation}
\label{eq-eigen-pp}
\det [ \hat I_{11}(1,2) \hat I_{11}(2,3) e^{-iq_2d} +
\hat I_{12}(1,2) \hat I_{21}(2,3) e^{iq_2d} ] = 0.
\end{equation}

In what follows, we provide the explicit equations satisfied by the dispersion relation of the
surface waves in the three cases considered in Appendix C.

\emph{Case 1:} In this case, Eq.~(\ref{eq-eigen-pp}) leads to
\begin{equation}
\label{dr-case1}
e^{-iq_2d} = \pm \left( \frac{q_{\rm e} - \epsilon_{xx} q_2}
{q_{\rm e} + \epsilon_{xx} q_2} \right) , 
\end{equation}
where $q_{\rm e}$ is given in Eq.~(\ref{eq-qs-case1}). This equations reduces to 
Eq.~(\ref{dr-Hz}) in the electrostatic limit $k \gg \omega/c$.

\emph{Case 2:} Here, assuming that the surface wave propagates along
the $x$-direction, its dispersion relation satisfies the following relation
\begin{equation}
\label{dr-case2}
e^{-iq_2d} = \pm \left( \frac{q_{\rm e} - \epsilon_{xx} q_2}
{q_{\rm e} + \epsilon_{xx} q_2} \right) ,
\end{equation}
where $q_{\rm e}$ is given in Eq.~(\ref{eq-qs-case2}). In the electrostatic limit,
this equation reduces to Eq.~(\ref{dr-Hz}).

\emph{Case 3:} In this case, and assuming that the surface waves propagate along the 
$y$-direction, its dispersion relation is given by the solution of the following equation 
\begin{equation}
\label{dr-case3}
e^{-2iq_2 d} = \frac{(\eta_{yy}q_{\rm o,2}-q_2+\eta_{yz}k) (\eta_{yy}q_{\rm o,2}-q_2-\eta_{yz}k)}
{(\eta_{yy}q_{\rm o,2}+q_2+\eta_{yz}k) (\eta_{yy}q_{\rm o,2}+q_2-\eta_{yz}k)},
\end{equation}
where $\eta_{yy}=\epsilon_{yy}/(\epsilon_{yy}^2+\epsilon_{yz}^2)$,
$\eta_{yz}=-\epsilon_{yz}/(\epsilon_{yy}^2+\epsilon_{yz}^2)$, and $q_{\rm o,2}$ is given in 
Eq.~(\ref{eq-qs-case3}). In the electrostatic limit this equation reduces to Eq.~(\ref{dr-Hx}).


\end{document}